\documentclass{article}

\usepackage{arxiv}

\usepackage[utf8]{inputenc} % allow utf-8 input
\usepackage[T1]{fontenc}    % use 8-bit T1 fonts
\usepackage{hyperref}       % hyperlinks
\usepackage{url}            % simple URL typesetting
\usepackage{booktabs}       % professional-quality tables
\usepackage{amsfonts}       % blackboard math symbols
\usepackage{nicefrac}       % compact symbols for 1/2, etc.
\usepackage{microtype}      % microtypography
\usepackage{lipsum}		% Can be removed after putting your text content
\usepackage{graphicx}
\usepackage{subcaption}
\usepackage{doi}

\usepackage{amssymb, amsmath, amsthm}
\usepackage{pdfpages}
\usepackage{tcolorbox}
\usepackage{enumitem}
\usepackage{tikz}
\usepackage{float}
\usepackage{pgfplots}
\usepackage{cite}

\DeclareFontFamily{OMS}{oasy}{\skewchar\font48 }
\DeclareFontShape{OMS}{oasy}{m}{n}{%
         <-5.5> oasy5     <5.5-6.5> oasy6
      <6.5-7.5> oasy7     <7.5-8.5> oasy8
      <8.5-9.5> oasy9     <9.5->  oasy10
      }{}
\DeclareFontShape{OMS}{oasy}{b}{n}{%
       <-6> oabsy5
      <6-8> oabsy7
      <8->  oabsy10
      }{}
\DeclareSymbolFont{oasy}{OMS}{oasy}{m}{n}
\SetSymbolFont{oasy}{bold}{OMS}{oasy}{b}{n}

\DeclareMathSymbol{\smallleftarrow}     {\mathrel}{oasy}{"20}
\DeclareMathSymbol{\smallrightarrow}    {\mathrel}{oasy}{"21}
\DeclareMathSymbol{\smallleftrightarrow}{\mathrel}{oasy}{"24}

\title{Anomalous Klein tunnelling with magnetic barriers in strained graphene}
\author{ \href{https://orcid.org/0009-0001-0666-7207}{\includegraphics[scale=0.06]{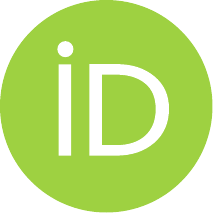}}\hspace{1mm}Edgardo Marin-Colli$^{\dagger,1}$,  \href{https://orcid.org/0009-0006-8696-5859}{\includegraphics[scale=0.06]{orcid.pdf}}\hspace{1mm}Tonatiuh Gómez-Ramírez$^{\dagger,2}$,  \href{https://orcid.org/0009-0003-3116-2088}{\includegraphics[scale=0.06]{orcid.pdf}}\hspace{1mm}O-Excell Gutierrez$^{*,3}$,\\
{\bf \href{https://orcid.org/0000-0002-2941-6048}{\includegraphics[scale=0.06]{orcid.pdf}}\hspace{1mm}Yonatan Betancur-Ocampo$^{\star,4}$, \href{https://orcid.org/0000-0002-5394-8634}{\includegraphics[scale=0.06]{orcid.pdf}}\hspace{1mm}Alfredo Raya$^{*\ast,*\ast\ast,5}$ and \href{https://orcid.org/0000-0002-2180-3895}{\includegraphics[scale=0.06]{orcid.pdf}}\hspace{1mm}Erik Díaz-Bautista$^{\dagger,6}$}\\
$^{\dagger}$Escuela Superior de Física y Matemáticas, Instituto Politécnico Nacional, 07738 Mexico City, Mexico\\
$^{*}$Facultad de Ciencias, Universidad Autónoma de Baja California, 22860, Ensenada, Baja California, México\\
$^{\star}$Instituto de Física, Universidad Nacional Autónoma de México, Ciudad de México, México\\
$^{*\ast}$Facultad de Ingeniería Eléctrica, Universidad Michoacana de San Nicolás de Hidalgo, 58040 Morelia, Michoacán, México\\
$^{*\ast\ast}$Centro de Ciencias Exactas, Universidad del Bío-Bío. Casilla 447, Chillán, Chile\\
e-mail:  emarinc2200@alumno.ipn.mx$^1$, agomezr2201@alumno.ipn.mx$^2$, oscar.gutierrez22@uabc.edu.mx$^3$,\\ ybetancur@fisica.unam.mx$^4$, alfredo.raya@umich.mx$^5$, ediazba@ipn.mx$^6$}

\date{} 					% Or removing it

\renewcommand{\shorttitle}{Anomalous Klein tunnelling with magnetic barriers in strained graphene}

\hypersetup{
pdftitle={Electron transmission through magnetic barriers in strained graphene},
pdfsubject={mathematical physics, quantum mechanics, two-dimensional Dirac materials},
pdfauthor={Edgardo Marin-Colli, Tonatiuh Gómez-Ramírez, Oscar Excell Gutierrez Rodriguez, Erik Díaz-Bautista},
pdfkeywords={electron transmission, transfer matrix, strained graphene, magnetic and electrostatic barriers},
}

\begin{document}
\maketitle

\begin{abstract}
	We study electron transport in a strained graphene sheet subjected to a sequence of 
$N$ electrostatic and magnetic barriers. Employing a modified and improved transfer-matrix framework, we examine how the transmission and reflection coefficients evolve with variations in uniaxial strain and in the number of barriers. The interplay of mechanical deformation and external magnetic fields is found to generate an anomalous Klein tunnelling, allowing the conductance to be effectively modulated through strain and barrier configurations. These findings highlight the role of strain engineering and magnetic field modulation as powerful tools for tailoring charge transport in two-dimensional materials. More broadly, they underscore how mechanical and electromagnetic control can be used to design next-generation solid-state devices with tunable electronic properties.
\end{abstract}

% keywords can be removed

\section{Introduction}\label{sec1}

With the discovery of graphene, it became possible to explore Klein tunnelling experimentally — a relativistic effect originally predicted by Oskar Klein in 1929 \cite{Klein1929} that consists in a relativistic particle to tunnel with certainty regardless of the height and the width of the barrier. In graphene, charge carriers behave as massless Dirac fermions moving at a Fermi velocity $v_{\rm F}$ nearly 300 times slower than the speed of light \cite{Novoselov2005}. This property, together with the conservation of pseudo-spin, allowed perfect electron transmission through electrostatic barriers, even when the potential height exceeds the carrier energy \cite{Katsnelson2006,Allain2011,Young2009}. The conservation of pseudo-spin forbids backscattering at normal incidence, leading to a transmission coefficient $T = 1$ \cite{Katsnelson2006}.

The experimental realization of Klein tunnelling in graphene p-n junctions marked a transformational moment in condensed matter physics \cite{Young2009,Stander2009}. These experiments provided direct evidence of perfect transmission through engineered potential barriers, confirming theoretical predictions about the chiral nature of graphene quasiparticles. Subsequent work demonstrated conductance oscillations in narrow graphene heterojunctions, where the phase shift observed in the conductance fringes at low magnetic fields constituted a signature of the absence of backscattering at normally incident carriers \cite{Young2009}. These findings established graphene as an accessible platform for studying quantum electrodynamic phenomena that would be impractical to observe with elementary particles.

Beyond this ideal scenario, more complex configurations such as magnetic fields, strain, or periodic potentials can alter the transmission properties, giving rise to anomalous Klein tunnelling, where perfect transmission occurs at nonzero incidence angles or specific energy ranges \cite{Robinson2012,Yokoyama2008}. Strain engineering, in particular, has emerged as a powerful tool for modulating the electronic properties of graphene. Uniaxial strain can open a bandgap at the Dirac points, with tensile strains that can open band gap beyond 23\% along the zig-zag direction \cite{Pereira}, while combinations of shear and uniaxial strain can achieve bandgap modulations from 0 to 0.9 eV under deformations of 12-17\% \cite{Guinea2010,Shao2019}. The interplay between mechanical deformation and external magnetic fields introduces additional degrees of freedom for controlling transport properties, enabling the emergence of valley-dependent phenomena and pseudomagnetic fields that further enrich the physics of electron transmission \cite{Guinea2010,Lu2022}.

Graphene superlattices, consisting of periodic arrays of electrostatic and magnetic barriers, have attracted considerable attention as platforms for investigating transmission resonances and filtering properties \cite{Park2008,BarbierPeeters2009,RodriguezVargas2019}. The transfer matrix method has proven to be an essential theoretical tool for analyzing transport through these multi-barrier systems, allowing for the systematic study of how structural parameters such as barrier height, width, and periodicity affect the transmission spectrum \cite{RodriguezVargas2019,BarbierPeeters2010}. Non-conventional superlattice profiles, including Gaussian and Lorentzian potential distributions, have demonstrated the ability to create nearly perfect pass bands and omnidirectional filtering characteristics \cite{RodriguezVargas2019}. These findings suggest potential applications in electron beam collimation, energy filtering, and valleytronic devices.

In this work, we study this phenomenon using a graphene superlattice model \cite{Park2008,BarbierPeeters2009,BarbierPeeters2010,GhoshSharma2009,Yankowitz2012,Ponomarenko2013,Dean2013} composed of $N$ electrostatic and $\delta$-magnetic barriers arranged along the $x$-axis. This approach enables us to investigate how mechanical deformation and field modulation affect the transmission resonances ($T = 1$), providing a framework to control and interpret both Klein and anomalous Klein tunnelling in graphene-based systems. The prospect of manipulating electron transport through the combined effects of strain and $\delta$-magnetic fields opens new pathways toward graphene-based electronic devices, including high-frequency transistors, quantum computing platforms, and flexible electronics \cite{DEHEER200792,Lin2010,Schwierz2010}. We organize the remaining of this article as follows: In Sect.~\ref{sec2} we construct the effective Hamiltonian for uniaxially strained graphene. Section~\ref{sec3} is devoted to the discussion of the transfer matrix framework we employ in our study. In Sect.~\ref{sec4} we present and discuss the results for the transmission coefficient $T$ and conductance $G$ for different barrier configurations. Conclusions are drawn at the end, in Sect.~\ref{conclusions}.

\section{Effective Dirac Hamiltonian for uniaxially strained graphene}\label{sec2}
\begin{figure}[ht]
	\centering
	\includegraphics[width=0.85\linewidth]{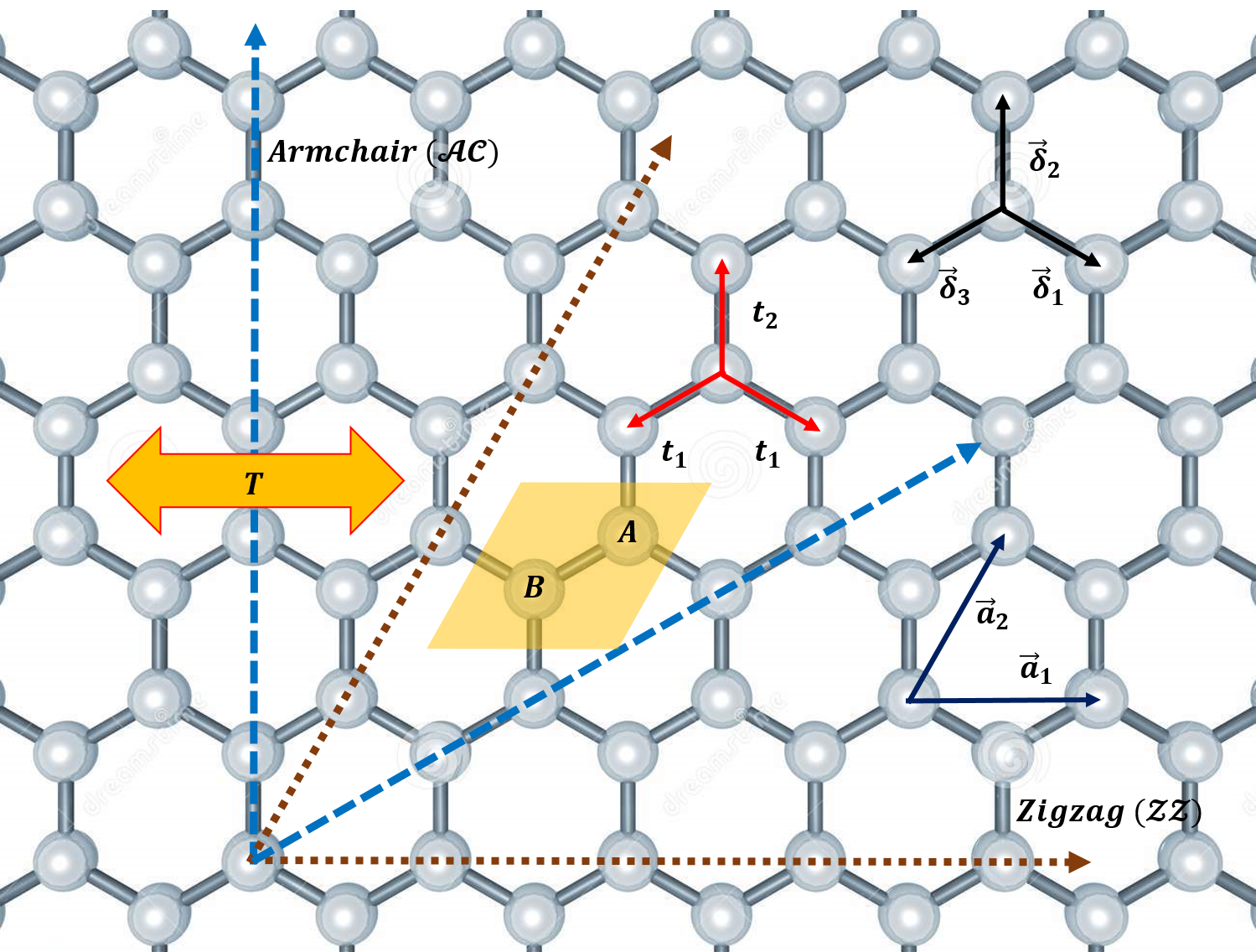}
	\caption{Uniaxially strained graphene is considered with the zigzag (armchair) direction aligned along the $x$-axis ($y$-axis), alternating every $30^\circ$. Strain along these high-symmetry directions modifies the nearest-neighbor hopping amplitudes to two distinct values, $t_{1}$ and $t_{2}$. The nearest-neighbor positions are given by $\vec{\delta}_{1}$, $\vec{\delta}_{2}$, and $\vec{\delta}_{3}$, while the lattice vectors $\vec{a}_{1}$ and $\vec{a}_{2}$ span the deformed hexagonal structure. The unit cell is colored in yellow.}
	\label{fig:lattice}
\end{figure}

Graphene is a two-dimensional Dirac material well-known in condensed matter physics due to its unique electronic, optical and mechanical properties. It consists of a honeycomb-like lattice composed of carbon atoms and is described by the lattice vectors $\vec{a}_{1}$ and $\vec{a}_{2}$ shown in Fig.~\ref{fig:lattice}. The graphene unit cell contains two carbon atoms, each of which belongs to a triangular sublattice, labelled as A or B. In pristine graphene, as well as strained graphene~\cite{Pereira,og13,Midtvedt,Betancur2}, the tight-binding (TB) Hamiltonian that describes the hop of an electron to the nearest atom~\cite{Betancur2,Goerbig,Goerbig2}, i.e., from an atom in the sublattice A (B) to another one in sublattice B (A), is obtained analizing the reciprocal lattice, where second-nearest neighbor interactions and $p_z$ orbital overlap are neglected. From the Bloch theorem, such a Hamiltonian turns out to be~\cite{Betancur2,Goerbig,wallace47}:
\begin{equation}
H_{{\rm TB}} = \sum^3_{j = 1}\left(\begin{array}{cc}
0 & t_j{\rm e}^{i\vec{k}\cdot\vec{\delta}_j}\\
t_j{\rm e}^{-i\vec{k}\cdot\vec{\delta}_j} & 0
\end{array}\right),
\label{H}
\end{equation}
being its energy spectrum
\begin{equation}\label{TB-energy}
    E_{\pm}=\lambda\left(\sum_{i=1}^{3}t_{i}^2+2\sum_{i<j}^{3}t_{i}t_{j}\cos\left(\vec{k}\cdot(\delta_{i}-\delta_{j})\right)\right)^{1/2},
\end{equation}
where $\lambda={\rm sgn}(E)$ denotes the conduction $(\lambda=+1)$ and valence ($\lambda=-1$) energy bands, $\vec{k}=(k_{x},k_{y})$ is the wave vector of electron, while $t_1$ and $t_2$ are two hopping parameters that quantify the probability amplitude that an electron hops to the nearest atom. In the strained graphene case, $t_{j}$ vary with uniaxial strain $\epsilon$ through an exponential dependence on the modified bond lengths $\delta_{j}$, $j=1,2,3$, namely, $t_j = t\,{\rm e}^{-\beta(\delta_j/a_{0} - 1)}$, where $\beta=2.6$ is the Gr\"uneisen constant, $t=2.7$ eV is the hopping in pristine graphene and $a_0=1.42$ \r{A} is the bond length in the undeformed sheet of the material~\cite{Pereira,Betancur2,Castro,Papas}. Here, the tensile strain $\epsilon$ quantifies the percentage of deformation~\cite{Pereira} and is proportional to the magnitude of a uniaxial tension $T$ applied to a graphene layer along zigzag ($\mathcal{ZZ}$) or armchair ($\mathcal{AC}$) direction (see again Fig.~\ref{fig:lattice}).

\begin{figure}[htbp]
     \centering
     \begin{subfigure}{0.45\linewidth}
         \centering
    \includegraphics[width=\linewidth]{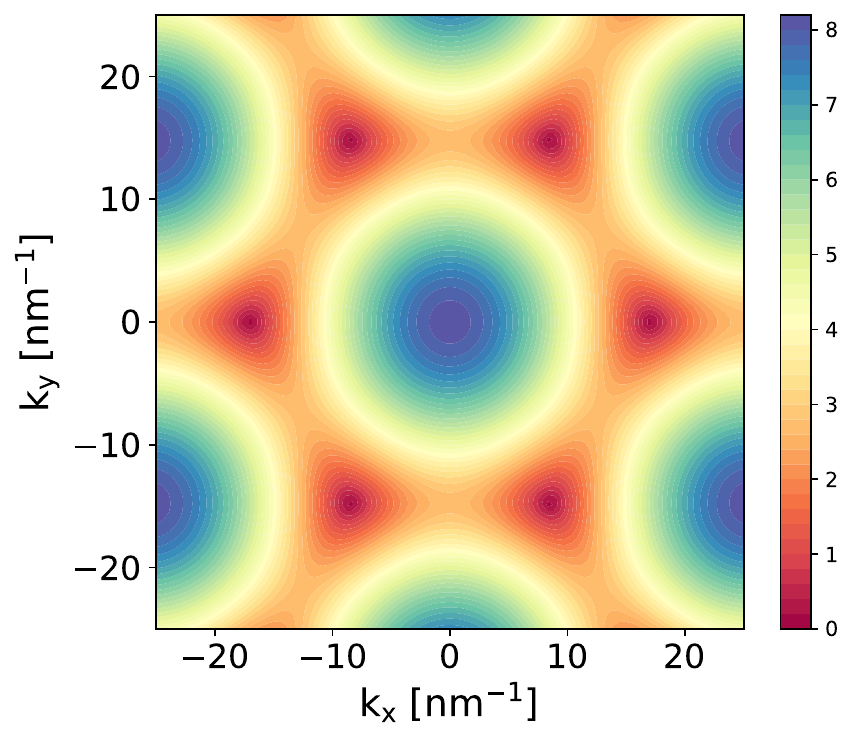}
        \caption{Pristine graphene ($\epsilon=0\%$)}
        \label{fig:2a}
     \end{subfigure}
     %\hfill
     \begin{subfigure}{0.45\linewidth}
         \centering
    \includegraphics[width=\linewidth]{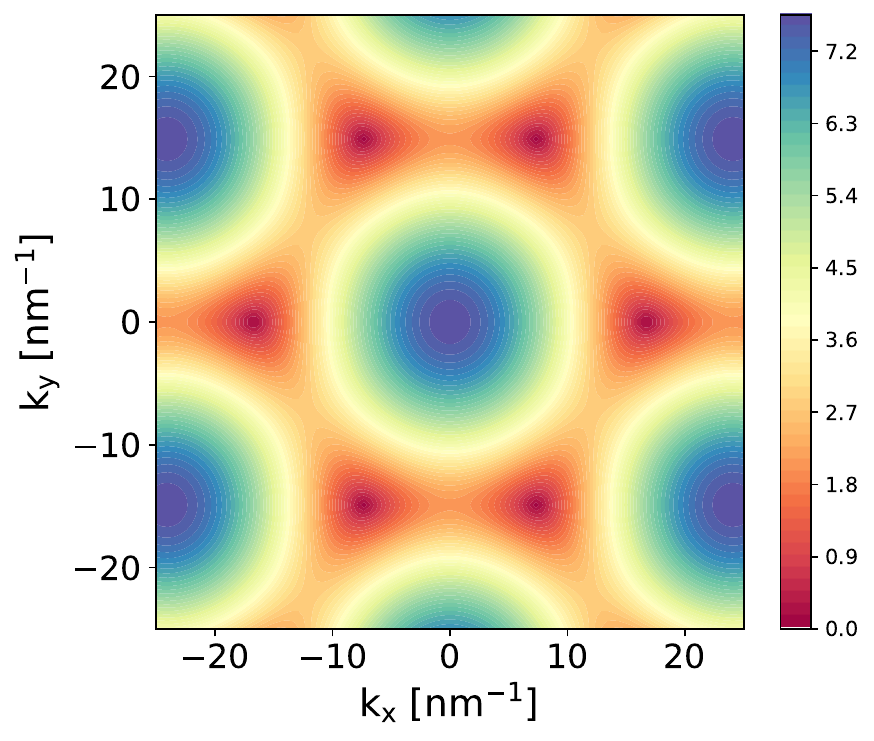}
        \caption{$\epsilon=6\%$ along the $\mathcal{ZZ}$ direction}
        \label{fig:2b}
     \end{subfigure}
     \begin{subfigure}{0.45\linewidth}
         \centering
        \includegraphics[width=\linewidth]{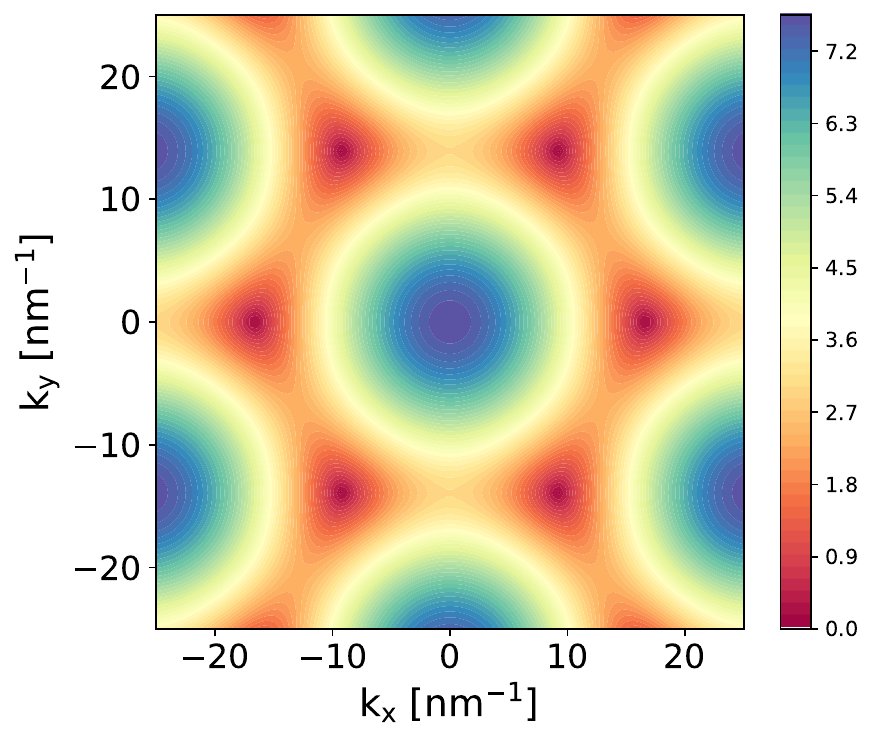}
        \caption{$\epsilon=6\%$ along the $\mathcal{AC}$ direction}
        \label{fig:2c}
     \end{subfigure}
     %\hfill
     \begin{subfigure}{0.45\linewidth}
         \centering
        \includegraphics[width=\linewidth]{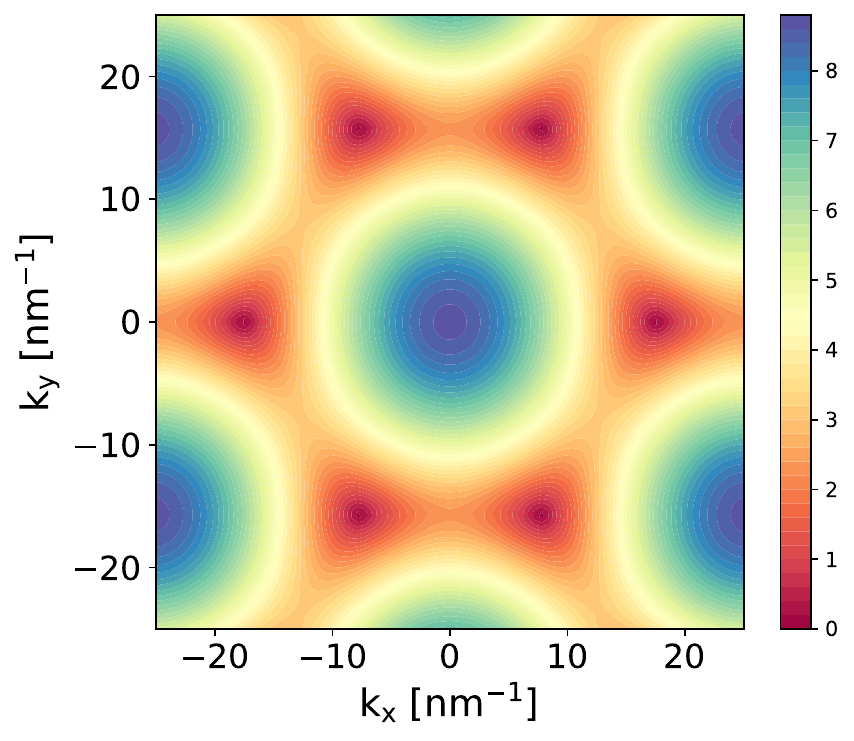}
        \caption{$\epsilon=-6\%$ along the $\mathcal{AC}$ direction}
        \label{fig:2d}
     \end{subfigure}
    \caption{Energy contour of the conduction band in \eqref{TB-energy} near the first Brillouin zone for different strain values $\epsilon$.}
    \label{fig:BrillouinZone}
\end{figure}

More precisely, the nearest-neighbor sites, $\vec{\delta}_1 = 2\vec{a}_1/3 - \vec{a}_2/3$, $\vec{\delta}_2 = 2\vec{a}_2/3 - \vec{a}_1/3$, and $\vec{\delta}_3 = -\vec{\delta}_1 -\vec{\delta}_2$, are related to the tensile strain through the deformed lattice vectors given by~\cite{diazbetancur20}: 
	\begin{align}\label{lattvszz}
	\vec{a}_1 &=  \sqrt{3}\, a_{0}\left[\,(1 + \rho^{+}\epsilon + \rho^{+}\epsilon \cos(2\zeta))\,{\bf e}_{x} 
\;+\; \rho^{+}\epsilon \sin(2\zeta)\,{\bf e}_{y}\,\right], \\
	\vec{a}_2 &=  \dfrac{\sqrt{3}}{2}\, a_{0}\left[\,(1 + \rho^{-}\epsilon + 2\rho^{+}\epsilon \cos(2\zeta - \pi/3))\,{\bf e}_{x} 
\;+\; \bigl(\sqrt{3}(1 + \rho^{-}\epsilon) + 2\rho^{+}\epsilon \sin(2\zeta - \pi/3)\bigr)\,{\bf e}_{y}\,\right],
	\end{align}
%for uniaxial strain in the $\mathcal{ZZ}$ direction:
%\begin{subequations}
%	\begin{equation}\label{lattvszz}
%	\vec{a}^{\mathcal{Z}}_1 =  \sqrt{3}a_0(1 + \epsilon){\bf e}_{x}, \quad
%	\vec{a}^{\mathcal{Z}}_2 =  \frac{\sqrt{3}}{2}a_0\left((1 + \epsilon){\bf e}_{x} + \sqrt{3}(1 -\nu\epsilon){\bf e}_{y}\right),
%	\end{equation}
%\end{subequations}
%or for the $\mathcal{AC}$ direction:
%\begin{subequations}
%	\begin{equation}\label{lattvsac}
%	\vec{a}^{\mathcal{A}}_1 =  \sqrt{3}a_0(1 - \nu\epsilon){\bf e}_{x}, \quad
%	\vec{a}^{\mathcal{A}}_2 =  \frac{\sqrt{3}}{2}a_0\left((1 - \nu\epsilon){\bf e}_{x} + \sqrt{3}(1 +\epsilon){\bf e}_{y}\right),
%	\end{equation}
%\end{subequations}
where $\rho^{\pm}=(1\pm\nu)/2$, being $\nu\sim0.165$ the Poisson ratio of graphene~\cite{Pereira2,Cadelano}, and $\zeta$ indicates the direction along the uniaxial strain $\epsilon$ is applied. Therefore, for $\mathcal{ZZ}$ deformations ($\zeta=0^\circ$), we have that
%\begin{subequations}
	\begin{equation}\label{dzz}
	\delta^{\mathcal{ZZ}}_1 =  \delta^{\mathcal{ZZ}}_3= a_0\sqrt{\left(1 + \frac{1}{4}(3 - \nu)\epsilon\right)^2 + \frac{3}{16}(1+ \nu)^2\epsilon^2}, \quad
	\delta^{\mathcal{ZZ}}_2 =  a_0(1 -\nu\epsilon),
	\end{equation}
%\end{subequations}
while for $\mathcal{AC}$ deformations ($\zeta=90^\circ$), we get
%\begin{subequations}
	\begin{equation}\label{dac}
	\delta^{\mathcal{AC}}_1 =  \delta^{\mathcal{AC}}_3= a_0\sqrt{\left(1 + \frac{1}{4}(1 - 3\nu)\epsilon\right)^2 + \frac{3}{16}(1+ \nu)^2\epsilon^2}, \quad
	\delta^{\mathcal{AC}}_2 =  a_0(1 + \epsilon).
	\end{equation}
%\end{subequations}

Now, expanding the TB Hamiltonian \eqref{H} around the Dirac point position $\vec{K}_{\rm D}$ at the first Brillouin zone (see Fig.~\ref{fig:BrillouinZone}), i.e. by taking $\vec{k} = \vec{p} + \vec{K}_{\rm D}$ such that $\vert\vec{p}\vert\ll\vert\vec{K}_{\rm D}\vert$, and where the following relation is fulfilled
\begin{equation}\label{hopping}
\sum^3_jt_j\exp\left(-i\vec{K}_{\rm D}\cdot\vec{\delta}_j\right) = 0 \quad \Longrightarrow \quad \cos[\vec{K}_{\rm D}\cdot(\vec{\delta}_1 - \vec{\delta}_2)] = -\frac{t_2}{2t_1},
\end{equation}
the effective Dirac-like Hamiltonian in the continuum approximation is given by
\begin{equation}\label{HD}
H=v_{x}\,p_{x}\sigma_{x}+v_{y}\,p_{y}\sigma_{y},
\end{equation}
where $\sigma_{x/y}$ are the Pauli matrices and
\begin{align}\label{ayb}
v_{x} &= \frac{1}{\hbar}\sum_{j=1}^{3}\delta_{jx}t_j\sin\left(\vec{K}_{\rm D}\cdot\vec{\delta}_{j}\right) =\frac{1}{\hbar}\sqrt{a^2_{1x}t^2_1 + (a_{2x} - a_{1x})a_{2x}t^2_2}, \\
v_{y} &= \frac{1}{\hbar}\sum_{j=1}^{3}\delta_{jx}t_j\cos\left(\vec{K}_{\rm D}\cdot\vec{\delta}_{j}\right)  =\frac{1}{\hbar}\sqrt{a^2_{1y}t^2_1 + (a_{2y} - a_{1y})a_{2y}t^2_2},
\end{align}
are the components of the Fermi velocity.

Then, from the Schr\"odinger equation
\begin{equation}
    H\Psi(\vec{r},\vec{k})=\hbar\begin{pmatrix}
        0 & g^{*}(\vec{k}) \\
        g(\vec{k}) & 0
    \end{pmatrix}\Psi(\vec{r},\vec{k})=E\Psi(\vec{r},\vec{k}),
\end{equation}
where $E$ denotes the energy and the function $g(\vec{k})=g_x(\vec{k})+ig_y(\vec{k})=v_xk_x+iv_yk_y$, we write the solution as:
\begin{equation}
    \Psi(\vec{r},\vec{k})=\frac{1}{\sqrt{2}}\begin{pmatrix}
        1\\ \lambda\,{\rm e}^{i \theta\left(\vec{k}\right)}
    \end{pmatrix}{\rm e}^{ik_xx}{\rm e}^{ik_yy},\quad E=\lambda\, \hbar\sqrt{v_x^{2}k_x^{2}+v_y^{2}k_y^{2}}.
\end{equation}
Here,
\begin{equation}
    \theta(\vec{k})=\arctan \left(g_y(\vec{k})/g_x(\vec{k})\right),
\end{equation}
is the so-called pseudo-spin angle, while wave vector direction is denoted by
\begin{equation}
    \gamma(\vec{k})=\arctan\left(k_y/k_x\right).
\end{equation}
It is worth mentioning that the group velocity direction, given by 
\begin{equation}\label{eq.15}
\phi(\vec{k})=\arctan\left(\partial_{k_y}\vert g(\vec{k})\vert/\partial_{k_x}\vert g(\vec{k})\vert\right),
\end{equation}
 indicates the direction of the electron beam in the scattering process and therefore, for anisotropic materials such as strained graphene \cite{Betancur2,díazbautista2022extended}, the angles $\theta(\vec{k})$, $\gamma(\vec{k})$, and $\phi(\vec{k})$ are not generally parallel, which have lead to the explanation of anomalous Klein tunnelling in strained graphene due to the conservation of pseudo-spin $\theta(\vec{k})$, while $\phi(\vec{k})$ follows an anisotropic Snell's law.

%It is important to mention that, for graphene under mechanical deformation, the pseudo-spin angle and the wave vector angle are not equal. $\theta(\vec{k})=\arctan\left(\frac{v_yk_y}{v_x k_x}\right)$
%In addition, the group velocity angle, given by the following expression:
%$\varphi(\vec{k})=\arctan \frac{\partial_{k_x} |g(\vec{k})|}{\partial_{k_y} |g(\vec{k})|}=\arctan \frac{v_y^{2}k_y}{v_x^{2}k_x}$
%Furthermore, the following relationship holds:$\tan \alpha=\frac{v_y^{2}}{v_x^2}\tan{\theta}$
{\color{black}It is worth mentioning that the effective Dirac Hamiltonian derived above is valid in the long-wavelength limit, where the electron wavelength is much larger than the lattice constant $a_{c}=\sqrt{3}\,a_{0} \sim 0.25$ nm and the spatial variations of external potentials occur on length scales significantly exceeding the interatomic distances. In this regime, the two inequivalent valleys $\vec{K}_{\rm D}$ and $\vec{K}_{\rm D}'$ in the first Brillouin zone remain decoupled, and intervalley scattering is negligible~\cite{Castro, Katsnelson2006}. %This approximation is valid when the characteristic length scale of the potential and magnetic field variations is much larger than $a_{c}$.
For these reasons, we shall assume that both the electrostatic potential $V(x)$ and the magnetic vector potential $A_y(x)$ considered in this work vary smoothly over distances of several nanometers. While we model magnetic barriers using Dirac delta functions for analytical tractability, these should be understood as representing sharp but finite-width transitions occurring over length scales $\sim 5{-}10$ nm, which is sufficient to prevent significant valley mixing~\cite{Ando2006, Peres2006}. In this way, the $\delta$-function idealization captures the essential physics of abrupt magnetic field changes while maintaining mathematical simplicity.

Similarly, the strain fields shall be assumed to vary smoothly across the sample, consistent with the continuum elasticity theory employed in Eqs.~\eqref{lattvszz}--\eqref{dac} \cite{Cao2020}. Atomic-scale disorder or defects, which could induce intervalley scattering, are not included in the present model. The results presented here thus apply to high-quality graphene samples where such disorder is minimal, as achieved in modern experiments on exfoliated or CVD-grown graphene~\cite{Dean2010, Wang2013}.

On the other hand, for potential variations on atomic length scales or in the presence of strong disorder, a full tight-binding treatment or atomistic simulations would be required to properly account for valley mixing effects~\cite{Leconte2011, Settnes}. Nevertheless, the continuum Dirac model employed here has been extensively validated against experiments and provides accurate predictions for transport properties in the regime where valley mixing is suppressed.}

\section{Transfer matrix framework}\label{sec3}
\begin{figure}[ht]
      \centering
      \includegraphics[width=\textwidth]{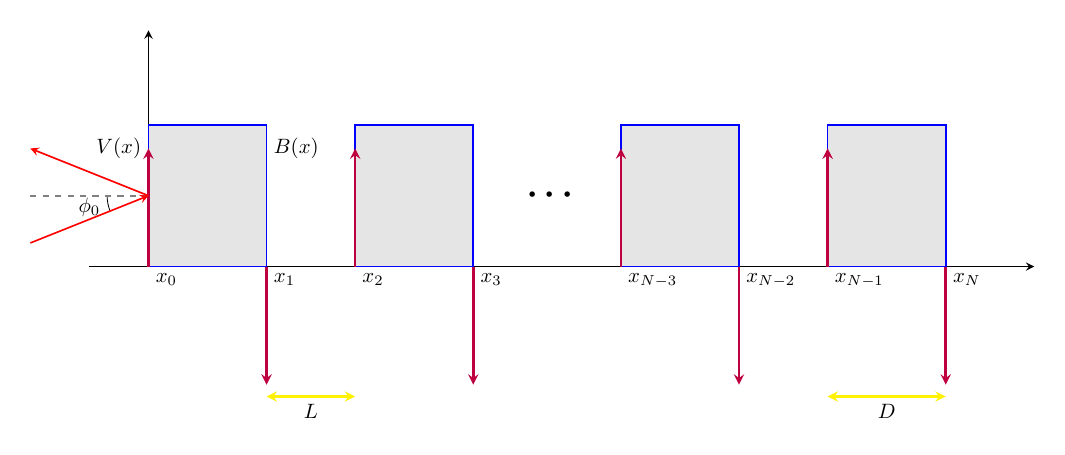}
      \caption{Diagram of the system of $N$ magnetic and electrostatic barriers disposed along the $x$ axis on a material surface.}
          \label{fig:Nmbarrier}
\end{figure}

Now, let us consider an electron in strained graphene under the interaction of a multiple-barrier structure of $N$ electrostatic potentials $V(x)$ and magnetic fields $\vec{B}(x)$. Assuming the Landau gauge $\vec{A}(x)=A_{y}(x){\bf e}_{y}$, the corresponding Hamiltonian is given by \cite{díazbautista2022extended,BEZERRA2021,Fernandes2023,FATTASSE2023,Belhadj2023}
      \begin{equation}
          H=\hbar \left(v_{x}\sigma_x p_x+v_{y}\sigma_y(p_y+e A_y(x))\right)+V(x)\sigma_0,
      \end{equation}
where
\begin{equation}\label{eq.17}
V(x)=\left\{\begin{array}{cc}
    V_{0}, & x_{2n-2}<x<x_{2n-1}, \\
    0, & \text{otherwise},
\end{array}\right.
 \qquad A_{y}(x)=B\ell_{B}\left(\Theta(x-x_{2n-2})-\Theta(x-x_{2n-1})\right),
\end{equation}
for $n=1,2,\dots,(N+1)/2$. Here, $D=x_{2n-1}-x_{2n-2}$ is the width of the $n$-th region, $N$ indicates the number of regions and $\ell_{B}=\sqrt{\hbar/(e\,B)}$ is the magnetic length (see Fig.~\ref{fig:Nmbarrier}). The function $\Theta(x)$ denotes the Heaviside step function, so that
\begin{equation}\label{eq.18}
    \vec{B}(x)=\nabla\times\vec{A}(x)=B\ell_{B}\left(\delta(x-x_{n-1})-\delta(x-x_{n})\right){\bf e}_{z},
\end{equation}
i.e., this corresponds to a pair of oppositely directed $\delta$-function magnetic barriers of strength $B$ and separated by a distance $D$. The choice of the vector potential $\vec{A}(x)$ as in \eqref{eq.17} guarantees that it is constant in each interval $x_{2n-1}<x<x_{2n-2}$. %{\color{magenta}It is important to recall that the low-energy charged carriers in graphene are described as Dirac fermions around the points $\vec{K}_{D}$ and $\vec{K}'_{D}$ of the Brillouin zone, separated by a momentum vector of the order $\vert \vec{K}_{D}-\vec{K}'_{D}\vert\sim1/a_{c}$, where $a_{c}=\sqrt{3}\,a_{0}=2.46$ \r{A} is the lattice constant. Therefore, in the presence of an external magnetic field, the vector potential $\vec{A}(x)$ would acquire rapid variations capable of inducing inter-valley couplings if $\ell_{B}\lesssim a_{c}$. Since in experimentally accessible magnetic fields the magnetic length is typically several orders of magnitude greater than the lattice constant, the valley mixture can be neglected in the graphene continuous model \cite{DEMARTINO2007}.

{\color{black}
As mentioned, the $\delta$-function magnetic barriers described in Eq.~\eqref{eq.18} constitute an idealization that facilitates the analytical and semi-analytical treatment of the transport problem. In reality, magnetic field profiles with vanishingly small width cannot be achieved, and any experimental realization would involve finite spatial extent. Nevertheless, this idealization serves a dual purpose: (i) it provides mathematical tractability, enabling us to derive exact or semi-exact solutions via the transfer matrix method, and (ii) it captures the essential physics of abrupt magnetic field transitions that can be engineered in graphene-based devices.

For the magnetic field strengths considered in this work ($B \sim 0.1{-}0.2$ T), the magnetic length $\ell_B = \sqrt{\hbar/(e\,B)}$ defines the characteristic scale for magnetic confinement. If the actual magnetic barrier width $D$ satisfies $D \ll \ell_B$, the $\delta$-function approximation provides a reasonable first-order description of the transport physics. However, for barrier widths comparable to or exceeding $\ell_B$, more complex effects emerge. In particular, the formation of Landau levels within the barrier regions becomes relevant, and the electronic states acquire characteristics of quantum dot confinement with a discretized energy spectrum~\cite{DEMARTINO2007}.

Then, to implement the transfer matrix approach, w}e propose the wave function {\it ansatz} inside the $n$-th region as
\begin{align}
    \Phi_n(\vec{r}) &= t_{n}\Psi(\vec{r},\vec{k}^t_n) + r_{n}\Psi(\vec{r},\vec{k}^r_n) \nonumber \\
    &=\frac{{\rm e}^{i k_y y}}{\sqrt{2}}\begin{pmatrix}
{\rm e}^{i k_{n,x}^t x} & {\rm e}^{i k_{n,x}^r x} \\
\lambda_{n} {\rm e}^{i (\theta(k_{n,x}^{t})+k_{n,x}^t x)} & \lambda_{n} {\rm e}^{i (\theta(k_{n,x}^{r})+k_{n,x}^r x)}
\end{pmatrix}\begin{pmatrix}
t_{n} \\ r_{n}
\end{pmatrix} 
    =\frac{{\rm e}^{i k_y y}}{\sqrt{2}}M_{n}(x)\begin{pmatrix}
t_{n} \\ r_{n}
\end{pmatrix},
\end{align}
where $t_{n}$ and $r_{n}$ are the transmitted and reflected wave amplitudes, respectively. For the first region ($n=0$), $t_0 = 1$ corresponds to the incident wave, with an incident wave vector $\vec{k}^t_0$, and $r_0 = r$ for the reflected part and wave vector $\vec{k}^r_0$. Notice that $n = N+1$ in the last region, and thus $r_{N+1} = 0$ because there is no reflection and $t_{N+1} = t$ for the outgoing wave. In addition, we define the dynamical quantities:
      \begin{align}
          k_{x,n}^{(t/r)}&=(+/-)\lambda_n \sqrt{\left(\frac{E-V(x)}{\hbar v_{x}}\right)^2-\frac{v_{y}^2}{v_{x}^2}\left(k_{y,0}+\frac{e A_{y}(x)}{\hbar}\right)^2}, \label{eq.19} \\
          k_{y,0}&=\frac{ v_{x} \vert E\vert \tan(\phi_{0})}{\hbar v_{y}^2 \sqrt{1+\left(\frac{v_{x}\tan(\phi_{0})}{v_{y}}\right)^2}}, \label{eq.20} \\
          \theta(\vec{k}^{(t/r)}_n)&=\arctan\left(\frac{v_{y}\left(k_{y,0}+\frac{e A_{y}(x)}{\hbar}\right)}{v_x k_{x,n}^{(t/r)}}\right),
      \end{align}
where $\phi_{0}\equiv\phi(\vec{k}^{t}_{0})$ is the incidence angle at the first barrier obtained from \eqref{eq.15}.

Now, by imposing the continuity condition at the position $x=x_{n}$ that separates the regions with wave functions $\Phi_n(\vec{r})$ and $\Phi_{n+1}(\vec{r})$,
    \begin{equation}
\Phi_{n}(x_{n})=\Phi_{n+1}(x_{n}),
    \end{equation}
we obtain the following relation
\begin{equation*}
    \left(\begin{array}{c}
          t_{n}\\
          r_{n}
    \end{array}\right) = M^{-1}_{n}(x_{n}) M_{n+1}(x_{n}) \left(\begin{array}{c}
          t_{n+1}\\
          r_{n+1}
    \end{array}\right),
\end{equation*}
where the matrices $M_{m}(x_{l})$ are
\begin{equation}
    M_{m}(x_{l}) = \frac{1}{\sqrt{2}}\left(\begin{array}{cc}
      {\rm e}^{ik^t_{x,m} x_l}  & {\rm e}^{ik^r_{x,m} x_l} \\
        \lambda_m {\rm e}^{i(\theta(\vec{k}^t_m)+k^t_{x,m} x_l)} & \lambda_m {\rm e}^{i(\theta(\vec{k}^r_m)+k^r_{x,m} x_l)}
    \end{array}\right).
\end{equation}

Now, considering the part of the system consisting of $N$ regions, it follows that
\begin{equation}
    \Phi_{0}(\vec{r})=M_{1}(x_{0})M^{-1}_{1}(x_{1}) M_{2}(x_{2})M^{-1}_{2}(x_{3}) \cdots M_{N}(x_{N-1})M^{-1}_{N}(x_{N}) \Phi_{N+1}(\vec{r}),
\end{equation}
from where we can define the effective transfer matrix as
\begin{equation}
\Lambda = \prod^N_{n=1} M_{n}(x_{n-1})M^{-1}_{n}(x_{n}),
\end{equation}
being
\begin{align}
    & M_{n}(x_{n-1})M^{-1}_{n}(x_{n}) \nonumber \\
    &\qquad =\frac{1}{{\rm e}^{i\theta(\vec{k}^r_n)}-{\rm e}^{i\theta(\vec{k}^t_n)}}
    \left(\begin{array}{cc}
        {\rm e}^{i(\theta(\vec{k}^r_n)-\xi^t_{n n-1})}-{\rm e}^{i(\theta(\vec{k}^t_n)-\xi^r_{n n-1})} & \lambda_n({\rm e}^{-i\xi^r_{n n-1}}-{\rm e}^{-i\xi^t_{n n-1}}) \\
        \lambda_n{\rm e}^{i(\theta(\vec{k}^t_n)+\theta(\vec{k}^r_n))}({\rm e}^{-i\xi^t_{n n-1}}-{\rm e}^{-i\xi^r_{n n-1}}) & {\rm e}^{i(\theta(\vec{k}^r_n)-\xi^r_{n n-1})}-{\rm e}^{i(\theta(\vec{k}^t_n)-\xi^t_{n n-1})}
    \end{array}\right),
\end{align}
%\end{widetext}
and $\xi^{(r/t)}_{n n-1} = k^{(r/t)}_{x,n}(x_n - x_{n-1})$ is the phase shift for the outgoing wave in the $n$-th region. The latter expression can also be expressed in terms of the propagation $P_{n}$ and dynamics $D_{n}$ matrices as
\begin{equation}
    M_{n}(x_{n-1})M^{-1}_{n}(x_{n})=D_{n}P_{n}D^{-1}_{n},
\end{equation}
where
\begin{equation}
    P_n=\begin{pmatrix}
        {\rm e}^{-ik^{t}_{x,n}(x_n - x_{n-1})} & 0 \\
        0 & {\rm e}^{-ik^{r}_{x,n}(x_n - x_{n-1})}
    \end{pmatrix}, \quad D_{n}=\begin{pmatrix}
        1 & 1 \\
        \lambda_{n}{\rm e}^{i\theta(\vec{k}^{t}_{n})} & \lambda_{n}{\rm e}^{i\theta(\vec{k}^{r}_{n})}
    \end{pmatrix}.
\end{equation}

The previous matrix transfer framework allows us to simplify the electron scattering across the stratified potential barriers, where it is only necessary to solve a $2 \times 2$ linear equation system to quantify the transmission probability of the electron crossing the whole set of barriers
      \begin{align}\label{eq.28}
    \Lambda\left(\begin{array}{c}
         1\\
         \lambda_{N+1}{\rm e}^{i\theta(\vec{k}^t_{N+1})}
    \end{array}\right){\rm e}^{ik^t_{x,N+1}x_N} t =\left(\begin{array}{c}
         1\\
         \lambda_0{\rm e}^{i\theta(\vec{k}^t_{0})}
    \end{array}\right) + r \left(\begin{array}{c}
         1\\
         \lambda_0{\rm e}^{i\theta(\vec{k}^r_{0})}
    \end{array}\right),
\end{align}
we find the squared modulus of transmission $t$ and reflection $r$ amplitudes as:
\begin{equation}\label{eq.29}
    T=|t|^2 = \frac{4\sin^2\left(\frac{1}{2}(\theta(\vec{k}^t_{0}) - \theta(\vec{k}^r_{0})\right)}{\vert\rho_y-\lambda_0\rho_x{\rm e}^{i\theta(\vec{k}^r_0)}\vert^2}, \quad R=\vert r\vert^{2}=\frac{|\rho_y-\lambda_0\rho_x{\rm e}^{i\theta(\vec{k}^t_0)}|^2}{|\rho_y -\lambda_0\rho_x{\rm e}^{i\theta(\vec{k}^r_0)}|^2},
\end{equation}
where
\begin{equation}\label{eq.30}
    \left(\begin{array}{c}
         \rho_x\\
         \rho_y 
    \end{array}\right) = \Lambda \left(\begin{array}{c}
        1\\
        \lambda_{N+1}{\rm e}^{i\theta(\vec{k}^t_{N+1})}
    \end{array}\right).
\end{equation}

\subsection{Conductance}
The conductance is a physical quantity defined as the inverse of resistance; therefore, its study provides important insights into the combined effects of the magnetic field and uniaxial strain on electronic transport in graphene. Based on the Landauer–B\"uttiker formalism, the conductance per unit length is calculated as~\cite{Landauer1957,Buttiker1985,Datta1995}
\begin{equation}\label{eq.32}
    G=\frac{e^2}{\hbar}\int T\,{\rm d}k=\frac{G_{0}}{\pi}\int_{-\pi/2}^{\pi/2} T\,\frac{\sec^2(\phi_{0})}{\left(1+\frac{v_{x}^2}{v_{y}^2}\tan^2(\phi_{0})\right)^{3/2}}\,{\rm d}\phi_{0},
\end{equation}
where
\begin{equation}\label{eq.33}
    G_{0}=\frac{e^2\vert E\vert v_{x}}{\hbar^2 v_{y}^2}.
\end{equation}
This formalism, which has become a cornerstone in mesoscopic physics, treats electron transport as a scattering problem at the Fermi level, where the conductance is determined by the transmission probability of electrons between reservoirs \cite{Datta1995,FoaTorres2014}. The approach is particularly suited for graphene systems, where phase-coherent transport can be maintained over micron-scale distances at low temperatures, and where the chiral nature of charge carriers leads to unique transmission properties \cite{Katsnelson2006,Tworzydlo2006}.

In the context of Klein tunnelling, the conductance exhibits distinctive features that directly reflect the anomalous transmission properties of Dirac fermions through potential barriers. Experimental measurements in graphene p-n junctions have demonstrated conductance oscillations that constitute direct signatures of Klein tunnelling \cite{Young2009,Stander2009}. These oscillations arise from Fabry-P\'erot resonances within the junction cavity, and their phase shift under low magnetic fields provides unambiguous evidence of the perfect transmission of normally incident carriers \cite{Young2009}. The resistance across steep potential steps has been shown to quantitatively agree with theoretical predictions of Klein tunnelling for Dirac fermions, confirming the validity of the Landauer-B\"uttiker description in this regime \cite{Stander2009,Huard2007}.

The application of external magnetic fields introduces additional complexity to the conductance behavior. Magnetic barriers deflect the angular dependence of electron transmission, suppressing normally incident electrons and reducing overall conductance \cite{Li2015,Yesilyurt2016}. When the magnetic field strength exceeds a critical value corresponding to the cyclotron orbit diameter matching the barrier width, conductance can be suppressed to nearly zero regardless of the potential barrier height, effectively achieving confinement of Dirac fermions \cite{Li2015}. This magnetic field-induced conductance modulation has been proposed as a mechanism for electro-optic applications and valleytronic devices \cite{Lu2022,Yesilyurt2016}. The interplay between magnetic barriers and electrostatic potentials generates distinct Fabry-P\'erot fringe patterns in the conductance, with characteristic constriction regions appearing when the barrier height approaches the Fermi energy \cite{Li2015}.

On the other hand, uniaxial strain provides an additional degree of freedom for conductance modulation in graphene systems. Strain engineering can open transport gaps at grain boundaries and interfaces, significantly affecting the on-off ratio in potential transistor applications \cite{Kumar2012,Nguyen2015}. The modulation of conductance depends sensitively on the topological structure of the boundaries, the strain direction, and the twist angle between misoriented graphene layers \cite{Kumar2012,Nguyen2015}. In vertical graphene heterostructures, strain-induced displacement of Dirac cones can create conduction gaps as large as a few hundred meV with only a few percent strain, leading to strong conductance modulation and significant enhancement of the Seebeck coefficient \cite{Nguyen2015}. Recent theoretical work predicts that strain-modulated graphene heterostructures can achieve on-off ratios up to $10^{12}$ while simultaneously enabling high current valley polarization, making them promising candidates for valleytronic current switches \cite{Chauwin2022}.

The combined effects of strain and magnetic fields on conductance in graphene superlattices offer promising pathways for device applications. Strain-induced pseudomagnetic fields can be tailored by pressure to achieve directionally selective electronic transmission and valley filtering, realizing basic elements for valleytronics \cite{Jones2017}. The ability to suppress Klein tunnelling through strain engineering addresses one of the fundamental challenges in graphene-based transistors, where the absence of backscattering typically prevents effective current switching \cite{Chauwin2022}. Conductance modulation through the interplay of electrostatic barriers, magnetic fields, and mechanical deformations thus provides a versatile framework for controlling electron transport in graphene, with implications for flexible electronics, strain sensors, and quantum computing platforms \cite{Schwierz2010,Kumar2012,Nguyen2015}.

%The conductance is a physical quantity defined as the inverse of resistance; therefore, its study provides important insights into the combined effects of the magnetic field and uniaxial strain on electronic transport in graphene. Based on the Landauer–Büttiker formalism, the conductance per unit length is calculated as:
%In pristine graphene....

\section{Results and discussion}~\label{sec4}
In this section, we analyze the case of a single magnetic and electrostatic barrier that allows us to obtain exact expressions with the possibility of corroboration of numerical calculations using the matrix transfer method from more than one barrier. Later, we consider the case of multiple barriers where the transmission is numerically calculated by using the matrix transfer method and Landauer-B\"uttiker formalism.

\subsection{Analytic case: a single magnetic and electrostatic barrier}
To obtain an explicit equation for the transmission and reflection coefficients, we will consider the case of a single magnetic and electrostatic barrier ($N=1)$ at $x_0=0$ and width $D$. 
%To illustrate clearly the combined effect of magnetic field strength and strain on transmission and reflection coefficients, we will consider the case of a single magnetic and electrostatic barrier ($N=1)$ at $x_0=0$ and width $D$.
From the left-hand side of \eqref{eq.30}, it follows
\begin{align}
    \rho_x&=({\rm e}^{i\theta(\vec{k}^r_1)}-{\rm e}^{i\theta(\vec{k}^t_1)})^{-1}  \left(( {\rm e}^{-i\xi^{t}_{10}}{\rm e}^{i\theta(\vec{k}^r_1)}-{\rm e}^{-i\xi^{r}_{10}}{\rm e}^{i\theta(\vec{k}^t_1)})+\lambda_{0}\lambda_{1}{\rm e}^{i\theta(\vec{k}^t_{0})}({\rm e}^{-i\xi^{r}_{10}}-{\rm e}^{-i\xi^{t}_{10}})\right), \\
    \rho_y&=({\rm e}^{i\theta(\vec{k}^r_1)}-{\rm e}^{i\theta(\vec{k}^t_1)})^{-1} \left(\lambda_{1}{\rm e}^{i\theta(\vec{k}^t_1)}{\rm e}^{i\theta(\vec{k}^r_1)}( {\rm e}^{-i\xi^{t}_{10}}-{\rm e}^{-i\xi^{r}_{10}})+\lambda_{0}{\rm e}^{i\theta(\vec{k}^t_{0})}( {\rm e}^{i\theta(\vec{k}^r_1)}{\rm e}^{i\xi^{r}_{10}}-{\rm e}^{i\theta(\vec{k}^t_1)}{\rm e}^{-i\xi^{t}_{10}})\right),
\end{align}
where we have made the identification $\xi^{(r/t)}_{10}=k_{x,1}^{(r/t)}D$, $\vec{k}^{t}_{2}=\vec{k}^{t}_{0}$ and $\lambda_{2}=\lambda_{0}$ for $x>x_{1}$.

Now, assuming the following approximation for the pseudo-spin angle of the transmitted and reflected waves:
\begin{equation}
    \theta(\vec{k}^{r}_{n})=\pi-\theta(\vec{k}^{t}_{n}), \quad n=0,1,
\end{equation}
and taking into account the sign change between the $x$ component of the wave vectors $\vec{k}^{(r/t)}_{n}$ in each region, we obtain:
\begin{align}
    T&= %\frac{\left(\cos(\theta(\vec{k}_{0}^{t}))\cos(\theta(\vec{k}_{1}^{t}))\right)^{2}}{\left(\cos(\theta(\vec{k}_{0}^{t}))\cos(\theta(\vec{k}_{1}^{t}))\cos(k_{x,1}^{t}D)\right)^{2}+\left(\sin(\theta(\vec{k}_{0}^{t}))\sin(\theta(\vec{k}_{1}^{t}))-\lambda_{0}\lambda_{1}\right)^{2}\sin^{2}(k_{x,1}^{t}D)} \nonumber \\&=
    \frac{\left(v_{x}^{2}k_{x,0}^{t}k_{x,1}^{t}\right)^{2}}{\left(v_{x}^{2}k_{x,0}^{t}k_{x,1}^{t}\cos(k_{x,1}^{t}D)\right)^{2}+\left(v_{y}^{2}k_{y,0}\left(k_{y,0}+\frac{e A_{y}(x)}{\hbar}\right)-\lambda_{0}\lambda_{1}\vert g(\vec{k}_{0}^{t})\vert\,\vert g(\vec{k}_{1}^{t})\vert\right)^{2}\sin^{2}(k_{x,1}^{t}D)}, \\
    R&= %\frac{\left(\sin(\theta(\vec{k}_{0}^{t}))-\lambda_{0}\lambda_{1}\sin(\theta(\vec{k}_{1}^{t}))\right)^{2}\sin^{2}(k_{x,1}^{t}D)}{\left(\cos(\theta(\vec{k}_{0}^{t}))\cos(\theta(\vec{k}_{1}^{t}))\cos(k_{x,1}^{t}D)\right)^{2}+\left(\sin(\theta(\vec{k}_{0}^{t}))\sin(\theta(\vec{k}_{1}^{t}))-\lambda_{0}\lambda_{1}\right)^{2}\sin^{2}(k_{x,1}^{t}D)}\nonumber \\    &=
    \frac{v_{y}^{2} \left(\vert g(\vec{k}_{1}^{t})\vert\, k_{y,0}-\lambda_{0}\lambda_{1}\vert g(\vec{k}_{0}^{t})\vert\,\left(k_{y,0}+\frac{e A_{y}(x)}{\hbar}\right)\right)^{2}\sin^{2}(k_{x,1}^{t}D)}{\left(v_{x}^{2}k_{x,0}^{t}k_{x,1}^{t}\cos(k_{x,1}^{t}D)\right)^{2}+\left(v_{y}^{2}k_{y,0}\left(k_{y,0}+\frac{e A_{y}(x)}{\hbar}\right)-\lambda_{0}\lambda_{1}\vert g(\vec{k}_{0}^{t})\vert\,\vert g(\vec{k}_{1}^{t})\vert\right)^{2}\sin^{2}(k_{x,1}^{t}D)}, \label{Coeff}
\end{align}
where $g(\vec{k}_{0}^{t})=v_{x}k_{x,0}^{t}+i v_{y}k_{y,0}$ and $g(\vec{k}_{1}^{t})=v_{x}k_{x,1}^{t}+i v_{y}\left(k_{y,0}+\frac{e A_{y}(x)}{\hbar}\right)$.
For pristine graphene, previous results coincide with those in \cite{Allain2011,Castro,Lejarreta_2013} when $B=0$ or in \cite{MYOUNG2009} for $V_{0}=0$, respectively. {\color{black}In Eq.~\eqref{Coeff}, it is possible to establish the condition for the anomalous Klein tunnelling angle, which is different from the Fabry-Pérot resonances. When $R = 0$, we find that
%\begin{equation}\label{KTanomalous}
%    \tan\phi_\textrm{KT} = -\frac{e\,v_yA_y(x)}{V_{0} \sqrt{1-\frac{e^2A^2_y(x)v^2_y}{V_{0}^2}}},
%\end{equation}
\begin{equation}\label{KTanomalous}
    \sin\phi_\textrm{KT} = -\frac{e\,v_yA_y(x)}{V_{0}},
\end{equation}
such an incidence angle $\phi_{0} = \phi_\textrm{KT}$ indicates the direction of perfect transmission, which also matches the refraction angle $\phi_t = \phi_0 = \phi_\textrm{KT}$ within the barrier. Therefore, this anomalous Klein tunnelling occurs when a ballistic electron inside the electrostatic and magnetic barriers keeps the beam trajectory straight.}

%\subsubsection{Graphics for a single barrier}
%In this section, we consider a system with a single potential barrier of height $V_0=14$ meV and width $\ell_B \approx 81. 1$ nm under the influence of a $\delta$-function magnetic barrier of strength $B=0.1$ T at the same region. The purpose of this section is to contrast the effect of uniaxial deformation applied along the $\mathcal{ZZ}$ and $\mathcal{AC}$ directions.

\begin{figure}[ht]
    \centering
    \begin{subfigure}{0.45\linewidth}
        \centering
         \includegraphics[width=\linewidth]{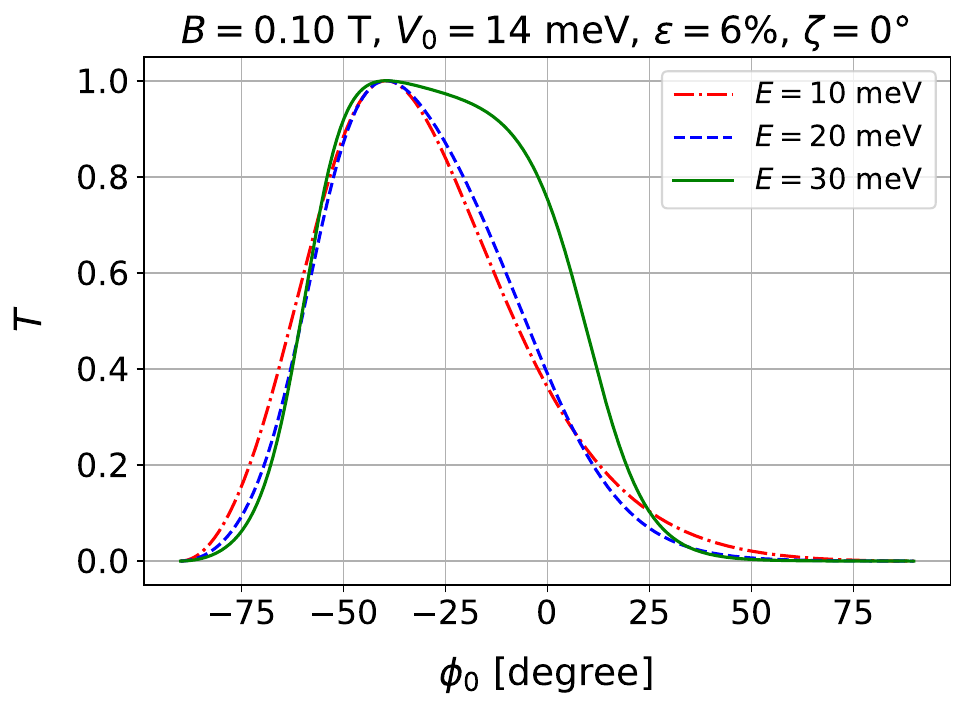}
         \caption{$\epsilon=6\%$ along the  $\mathcal{ZZ}$ direction}\label{fig:4a}
    \end{subfigure}
     \begin{subfigure}{0.45\linewidth}
        \centering
         \includegraphics[width=\linewidth]{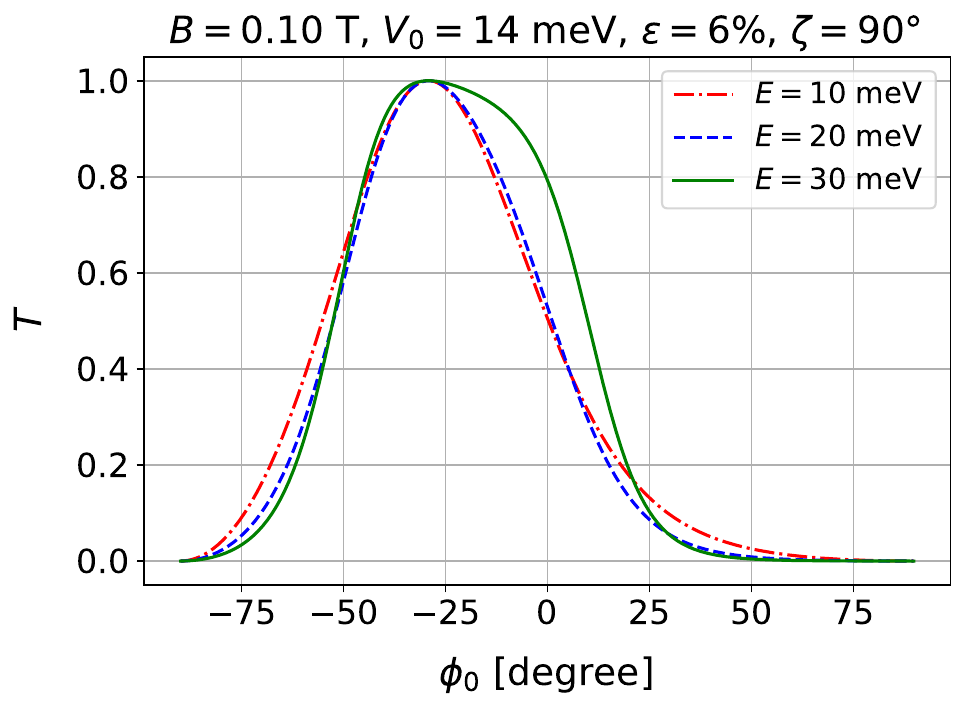}
         \caption{$\epsilon=6\%$ along the $\mathcal{AC}$ direction}\label{fig:4b}
    \end{subfigure}
     \begin{subfigure}{0.45\linewidth}
        \centering
         \includegraphics[width=\linewidth]{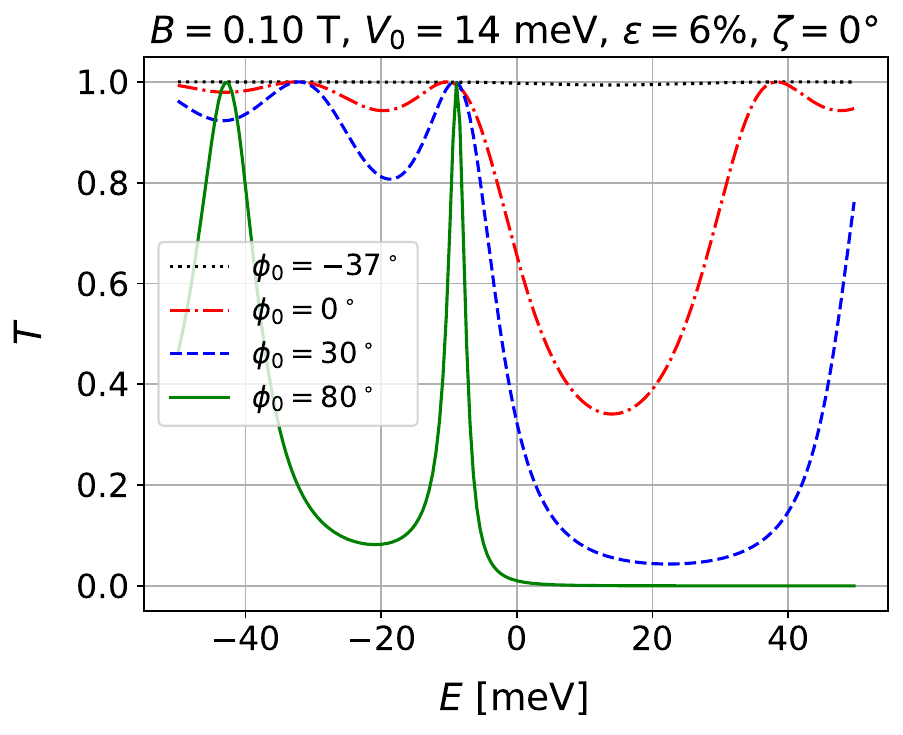}
         \caption{$\epsilon=6\%$ along the  $\mathcal{ZZ}$ direction}\label{fig:4c}
    \end{subfigure}
     \begin{subfigure}{0.45\linewidth}
        \centering
         \includegraphics[width=\linewidth]{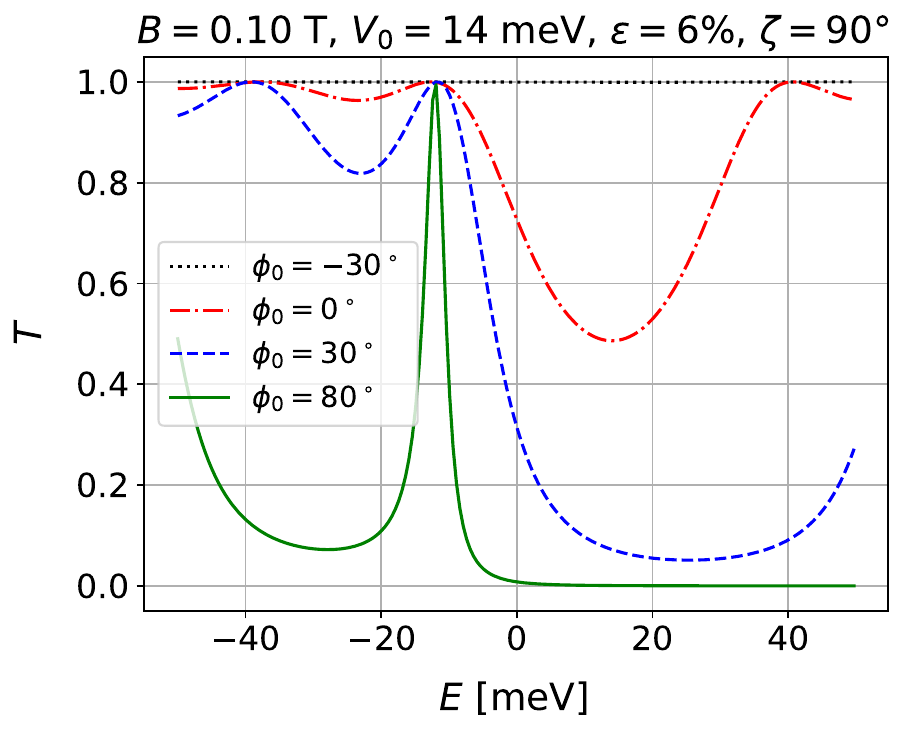}
         \caption{$\epsilon=6\%$ along the $\mathcal{AC}$ direction}\label{fig:4d}
    \end{subfigure}
    \caption{Electron transmission $T$ in uniaxially strained graphene with a single electrostatic and magnetic barrier. Panels (a) and (b) show the dependence on the incidence angle $\phi_{0}$ for different values of incident energy $E$. Panels (c) and (d) present the dependence on the Fermi energy $E$ for different incidence angles $\phi_{0}$. The electrostatic barrier height is fixed at $V_0 = 14$ meV, while the magnetic field strength is set to $B = 0.1$ T, with a width $D \approx 81.1$ nm for both. The black-dotted line in the lower graphs corresponds to anomalous Klein tunnelling.}
    \label{fig:Onebarrier}
\end{figure}

\begin{figure}[ht]
     \centering
     \begin{subfigure}{0.45\linewidth}
         \centering
    \includegraphics[width=\linewidth]{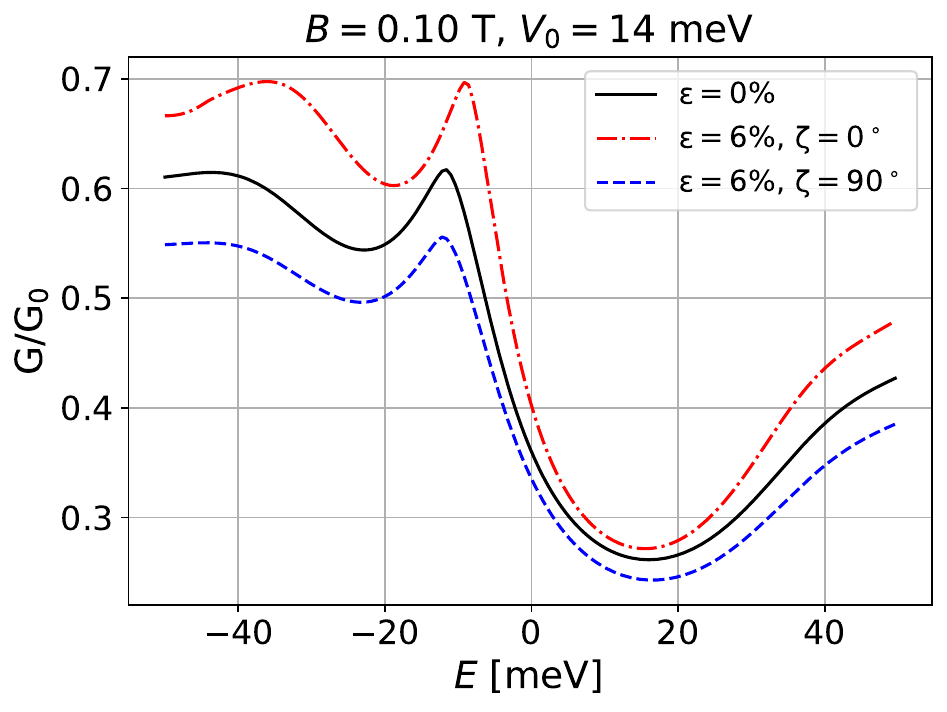}
        \caption{$B = 0.1$ T}
        \label{fig:5a}
     \end{subfigure}
     %\hfill
     \begin{subfigure}{0.45\linewidth}
         \centering
    \includegraphics[width=\linewidth]{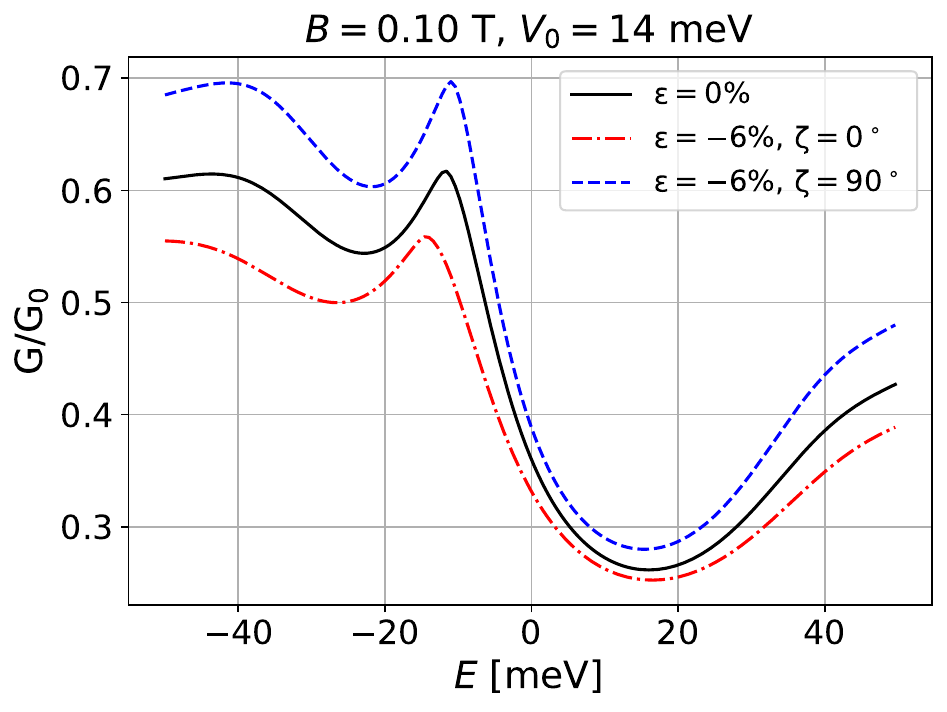}
        \caption{$B = 0.1$ T}
        \label{fig:5b}
     \end{subfigure}
     \begin{subfigure}{0.45\linewidth}
         \centering
        \includegraphics[width=\linewidth]{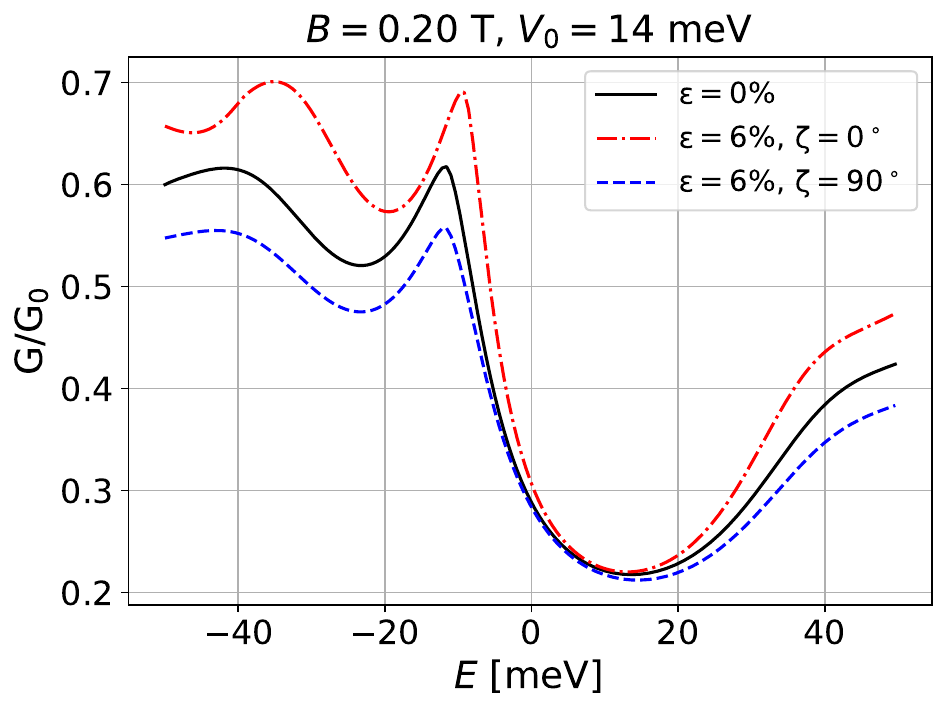}
        \caption{$B = 0.2$ T}
        \label{fig:5c}
     \end{subfigure}
     %\hfill
     \begin{subfigure}{0.45\linewidth}
         \centering
        \includegraphics[width=\linewidth]{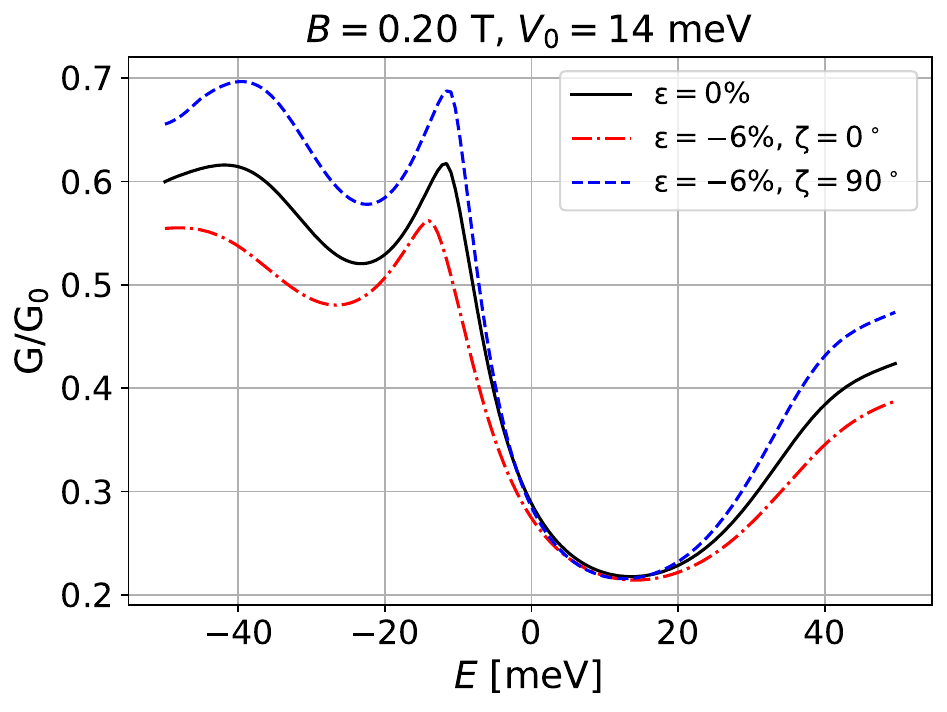}
        \caption{$B = 0.2$ T}
        \label{fig:5d}
     \end{subfigure}
    \caption{Conductance $G/G_{0}$ as a function of the Fermi level $E$ for a single electrostatic and magnetic barrier. The solid black, dashed red, and dotted blue curves correspond to pristine graphene and to strained graphene along the $\mathcal{ZZ}$ and $\mathcal{AC}$ directions, respectively. The parameters are set as $V_0 = 14$ meV and $D \approx 81.1$ nm.}
    \label{fig:1conductance}
\end{figure}

To contrast the effect in the coefficient transmission and conductance of uniaxial deformation applied along the $\mathcal{ZZ}$ and $\mathcal{AC}$ directions, let us consider a system with a single potential barrier of height $V_0=14$ meV and width $D=\ell_B \approx 81. 1$ nm under the influence of a $\delta$-function magnetic barrier of strength $B=0.1$ T at the same region (see Fig.~\ref{fig:Onebarrier}).% {\color{magenta}This selection of the magnetic field strength prevents valley mixing since $\ell_B\approx 329.67\,a_{c}$.}

As is well known, Klein tunelling is identified in the pristine case when perfect transmission ($T=1$) occurs for all values of incident energy $E$ at $\phi_0=0^\circ$ (normal incidence). However, if we introduce a mechanical deformation along the $\mathcal{ZZ}$ or $\mathcal{AC}$ direction, we observe a modification in such a tunnelling. This fact is particularly evident in panels (a) and (b) in Fig.~\ref{fig:Onebarrier}, where $T=1$ does not occur at zero angle. Furthermore, comparing with the pristine graphene case under the interaction of magnetic and electrostatic barriers \cite{MYOUNG2009}, the deformation along the $\mathcal{ZZ}$ direction is more susceptible to deviating from the pristine case than along the $\mathcal{AC}$ direction (see again Fig.~\ref{fig:Onebarrier}). 

Now, let us compare the panels (a) and (b) in Fig.~\ref{fig:Onebarrier} that correspond to two different configurations of mechanical deformations: along the $\mathcal{ZZ}$ and $\mathcal{AC}$ directions, respectively. These transmissions reveal that for {\color{black}$E = 10, 20, 30$ meV}, the incidence angle for which {\color{black}the perfect transmission occurs} in Fig.~\ref{fig:4b} lies closer to $\phi_0=0^\circ$ than in Fig.~\ref{fig:4a}. {\color{black}Actually, according to Eq.~\eqref{KTanomalous}, for Fig.~\ref{fig:4a}, $\phi_{\rm KT}\approx-36.04^{\circ}$ while for Fig.~\ref{fig:4b}, $\phi_{\rm KT}\approx-31.70^{\circ}$}. Moreover, for the incident energy $E = 30$ meV, the first one displays a large interval of angles $\phi_0$ for which the perfect transmission is possible, in clear contrast to Fig.~\ref{fig:4b}, whose transmission profile is almost similar to the graphs for $E = 10, 20$ meV.

%On the other hand, comparing panels (c) and (d), deformation along the $\mathcal{ZZ}$ axis produces more resonances than along the $\mathcal{AC}$ axis. This observation is consistent with theory, since the perfect transmission angle is now also influenced by the $v_x$ and $v_y$ parameters.%

On the other hand, by comparing panels (c) and (d) in Fig.~\ref{fig:Onebarrier} --also obtained for mechanical deformations applied to the $\mathcal{ZZ}$ ans $\mathcal{AC}$ directions, respectively-- we can see that for the incidence angle $\phi_0=0^\circ $, transmission coefficient reaches a local minimum value that lies between energies $E=10$ and $E=20$ meV in both panels (almost within the energy scale in the graph). {\color{black}However, the minimum transmission is about $T=0.3$ in Fig.~\ref{fig:4c} and around $T=0.5$ in Fig.~\ref{fig:4d}}. For $\phi_0=30^\circ$, the transmission coefficient in both figures starts to decrease from an energy close to {\color{black}$E =-10$ meV and almost reaches zero transmission near $E=20$ meV, but then in Fig.~\ref{fig:4c} it starts to increase more rapidly than in Fig.~\ref{fig:4d}}. Finally, for $\phi_0=80 ^\circ$, panel (c) exhibits two resonances, while panel (d) shows only one; in both, the transmission coefficient vanishes for some $E>0$ exhibiting minigaps in wider regions in this last panel, in contrast to the first one.

Overall, in Figs.~\ref{fig:4c} and \ref{fig:4d}, we observe that $T=1$ is reached at negative oblique angles as a consequence of resonances within the chosen energy range and the mechanical deformation applied. Moreover, for all three angles used, transmission coefficient in panel (c) decreases more sharply at positive Fermi energies than in panel (d). {\color{black}As we said before, the anomalous Klein tunnelling appears at the incidence angle $\phi_{\rm KT} \approx -36.04^\circ$ and $\phi_{\rm KT} \approx -31.70^\circ$} for the direction of $\mathcal{ZZ}$ and $\mathcal{AC}$ direction, respectively, which is unaffected by modulating the Fermi level. The occurrence of anomalous Klein {\color{black} tunnelling} is due to the conservation of pseudo-spin for these incidence angles, and it is only changed through the modulation of the magnetic field {\color{black}$B$ and the electrostatic potential $V_{0}$, as \eqref{KTanomalous} establishes}. As shown in Fig. \ref{fig:4c} and \ref{fig:4d}, other values of $T = 1$ are obtained by constructive interference inside the barrier giving rise to Fabry-Pérot resonances. Such resonances are controlled by changing the width or height barrier as well as the tuning of the Fermi level. %Furthermore, the regions where the transmission vanishes (well-known as minigaps \cite{Lejarreta_2013}) are wider in panel (d) than in (c) for the fixed angle $\phi_0 = 80^\circ $.

 %{\color{green}repeat} all three fixed angles, we observe that for positive Fermi energies $(E>0)$ in panel (c), transmission coefficient decreases rapidly at higher energies compared to panel (d). Moreover, for $\phi_0=0 ^\circ$ in (c), the minimums values of transmission coefficient is lower than in (d). Furthermore, the regions where the transmission is (well-known as minigaps \cite{Lejarreta_2013}) are wider in panel (d) than in (c) for the fixed angle $\phi_{0}=80 ^\circ $.
%

Now, let us compare the behavior of the conductance of Eq. \eqref{eq.32} in the pristine and strained graphene case. In Fig.~\ref{fig:5a}, strained graphene with $\epsilon= 6\%$ along the $\mathcal{ZZ}$ direction exhibits the highest conductance, whereas strain along the $\mathcal{AC}$ direction leads to the lowest values. The conductance of the pristine graphene lies between these two cases. In contrast, Fig.~\ref{fig:5b} shows the opposite behavior when a compress deformation is applied to the material: deformed graphene with $\epsilon= -6\%$ along the $\mathcal{AC}$ direction exhibits the highest conductance, whereas with $\epsilon= -6\%$ along the $\mathcal{ZZ}$ direction leads to the lowest values. This apparent interchange of the curves can be attributed to how the mechanical deformation modifies the crystalline structure, hopping parameters and, consequently, Fermi velocities in graphene. Then, when the magnetic field strength increases, the conductance is also modified as shown in Figs.~\ref{fig:5c} and~\ref{fig:5d}. More precisely, for a magnetic field strength $B=0.1$ T and for a given Fermi energy value $E$, the conductance obtained depends on the mechanical deformation applied and is always different between the pristine and strained graphene cases. However, when $B=0.2$ T, for instance, in the plots presented, resistance increases for $E>0$, as shown in Fig.~\ref{fig:5d}. Furthermore, there are some values of incident energy $E$ for which the conductance is the same in the pristine and strained graphene cases, as a consequence of the increase of $B$.

%In both panels, the conductance of the pristine graphene lies between these cases. The behavior of strained graphene with $\epsilon= 20\%$ along the $\mathcal{Z}=0 ^\circ$ in Fig.~\ref{fig:5a} is very similar with the behavior of strained graphene with $\epsilon= 20\%$ along the $\mathcal{Z}=90 ^\circ$ in Fig.~\ref{fig:5b}.%

\subsubsection{Fabry-Pérot resonances and critical angle}
\begin{figure}[htbp]
    \centering
    \begin{subfigure}{0.45\linewidth}
        \centering
         \includegraphics[width=\linewidth]{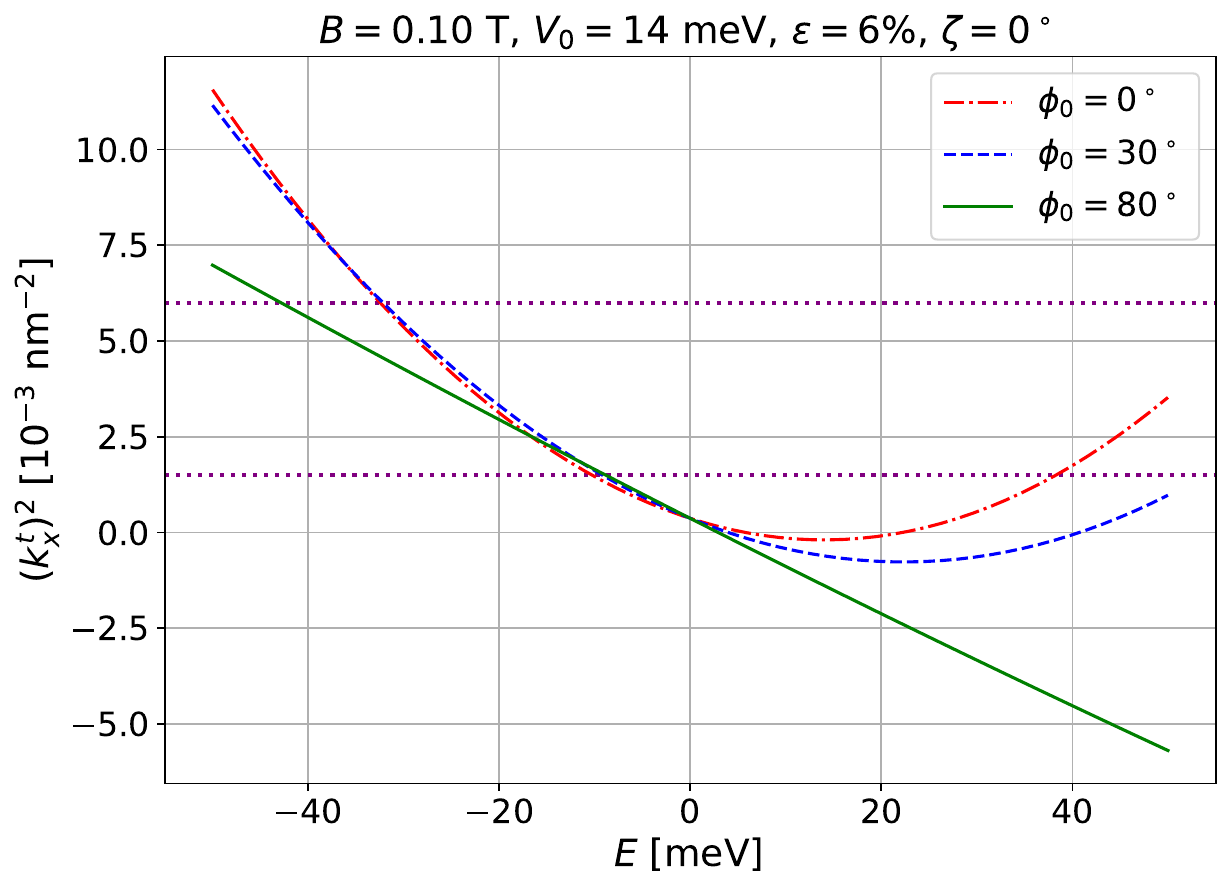}
         \caption{$\epsilon=6\%$ along the  $\mathcal{ZZ}$ direction}\label{fig:Resonance_a}
    \end{subfigure}
     \begin{subfigure}{0.45\linewidth}
        \centering
         \includegraphics[width=\linewidth]{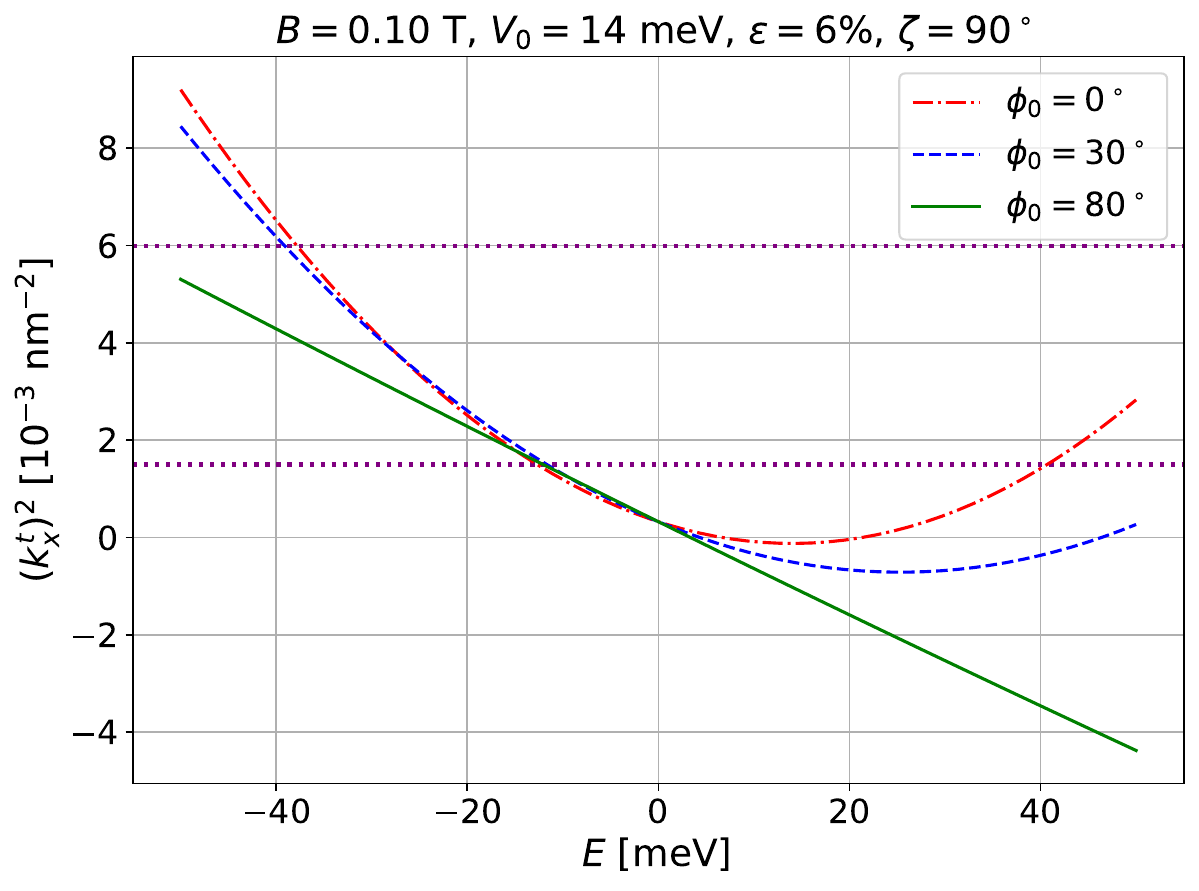}
         \caption{$\epsilon=6\%$ along the $\mathcal{AC}$ direction}\label{fig:Resonance_b}
    \end{subfigure}
    \caption{Behavior of the squared $x$-component of the transmitted wave number $\vec{k}_{1}^{t}$ in terms of the Fermi level $E$ for some incidence angles $\phi_0$ and different deformation configurations along the $\mathcal{ZZ}$ and $\mathcal{AC}$ directions. Horizontal purple dotted lines correspond to the values $n^2\pi^2/D^2$. tunnelling resonances are obtained when $k_{x,1}^{t}=n\,\pi/D$, as shown in Eq. \eqref{eq.40}.}
    \label{fig:Resonances}
\end{figure}

Delving deeper into the resonances that occur for oblique incidence ($\phi_0\neq0^\circ$), we recognize that a potential barrier can be analyzed as a double interface placed at $x=0$ and $x_1=D$, in analogy to a Fabry-Pérot interferometer \cite{Allain2011}. Therefore, if the incoming wave interferes constructively with itself between the two interfaces, transmission resonances will occur since the barrier will be transparent for electrons with $k_{x,1}^{t}=n\,\pi/D$ (tunnelling resonance). More explicitly, the condition of such resonances is written as:%which occurs when the energy $E$ and angle $\phi_{0}$ of incidence verify the relation:
\begin{equation}\label{eq.40}
    \left(\frac{n\,\pi }{D}\right)^{2}=\left(\frac{E-V_0}{\hbar\, v_{x}}\right)^2-\left(\frac{ \vert E\vert \tan(\phi_{0})}{\hbar\, v_{y} \sqrt{1+\left(\frac{v_{x}\tan(\phi_{0})}{v_{y}}\right)^2}}+\frac{e v_{y}A_{y}(x)}{\hbar\, v_{x}}\right)^2,
\end{equation}
according to \eqref{eq.19} and \eqref{eq.20} (see Fig.~\ref{fig:Resonances}).

\begin{figure}[htbp]
     \centering
     \begin{subfigure}{0.45\linewidth}
         \centering
    \includegraphics[width=\linewidth]{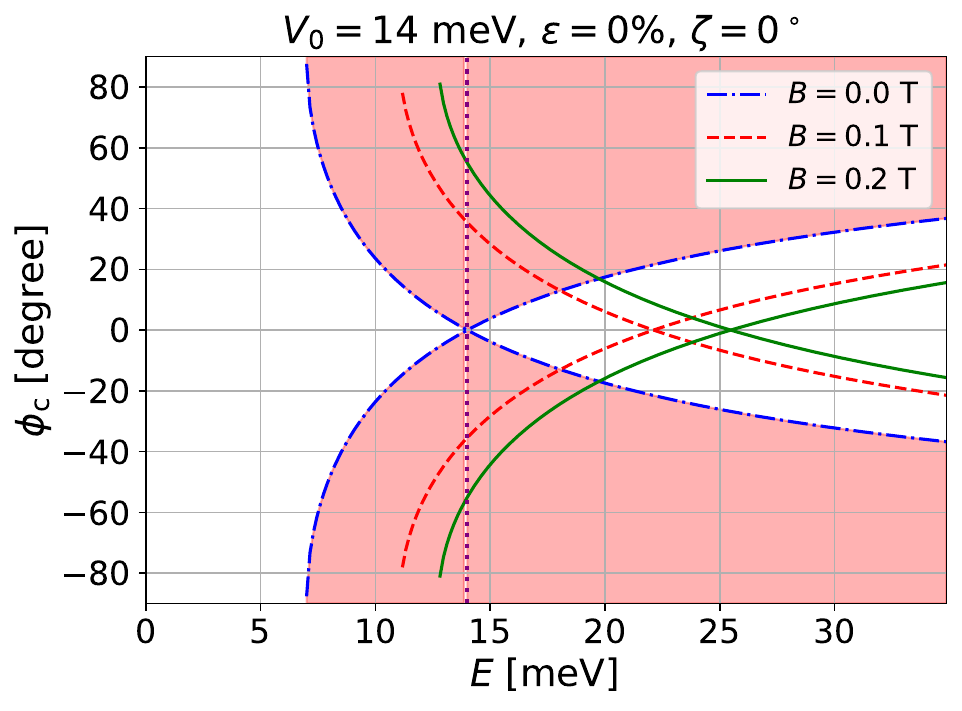}
        \caption{Pristine graphene ($\epsilon=0\%$)}
        \label{fig:3a}
     \end{subfigure}
     %\hfill
    %  \begin{subfigure}{0.45\linewidth}
    %      \centering
    % \includegraphics[width=\linewidth]{figures/angcritic_B02_eps00_zeta00_V14.pdf}
    %     \caption{Pristine graphene ($\epsilon=0\%$), $B=0.2$ T}
    %     \label{fig:3b}
    %  \end{subfigure}
     \begin{subfigure}{0.45\linewidth}
         \centering
        \includegraphics[width=\linewidth]{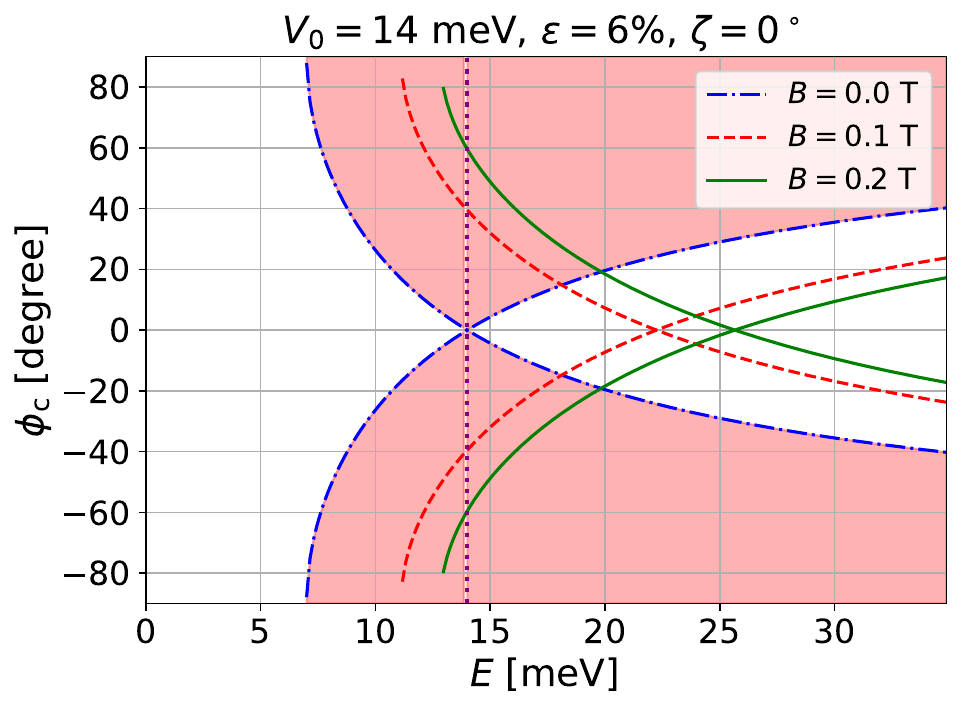}
        \caption{$\epsilon=6\%$ along the $\mathcal{ZZ}$ direction}
        \label{fig:3c}
     \end{subfigure}
     %\hfill
     % \begin{subfigure}{0.45\linewidth}
     %     \centering
     %    \includegraphics[width=\linewidth]{figures/angcritic_B02_eps20_zeta00_V14.pdf}
     %    \caption{$\epsilon=20\%$ along the $\mathcal{ZZ}$ direction, $B=0.2$ T}
     %    \label{fig:3d}
     % \end{subfigure}
          \begin{subfigure}{0.45\linewidth}
         \centering
        \includegraphics[width=\linewidth]{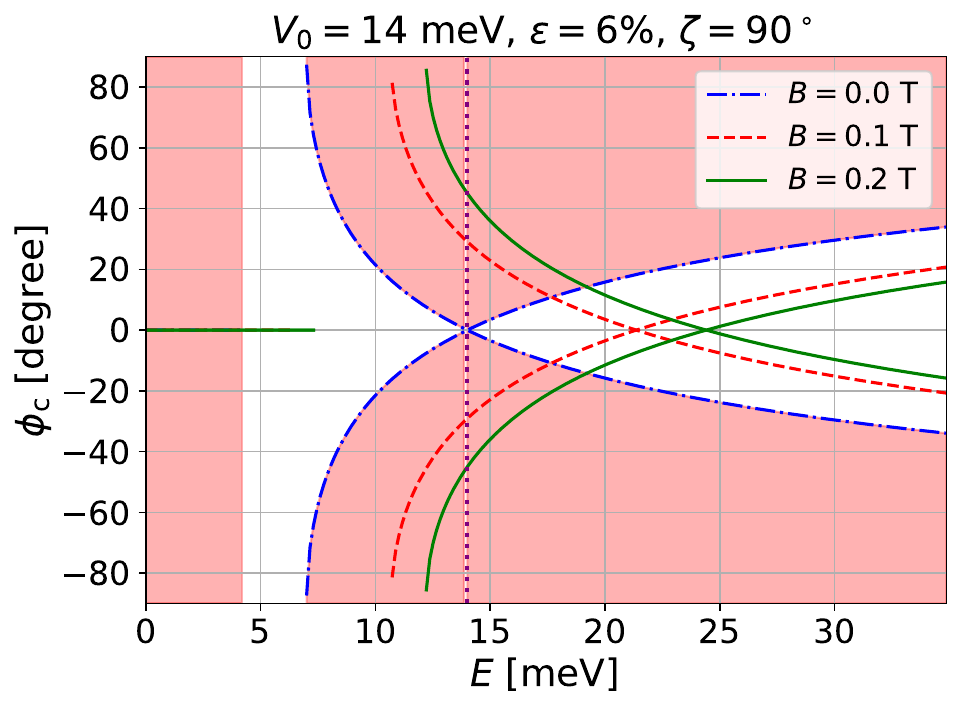}
        \caption{$\epsilon=6\%$ along the $\mathcal{AC}$ direction}
        \label{fig:3e}
     \end{subfigure}
     %\hfill
     \begin{subfigure}{0.45\linewidth}
         \centering
        \includegraphics[width=\linewidth]{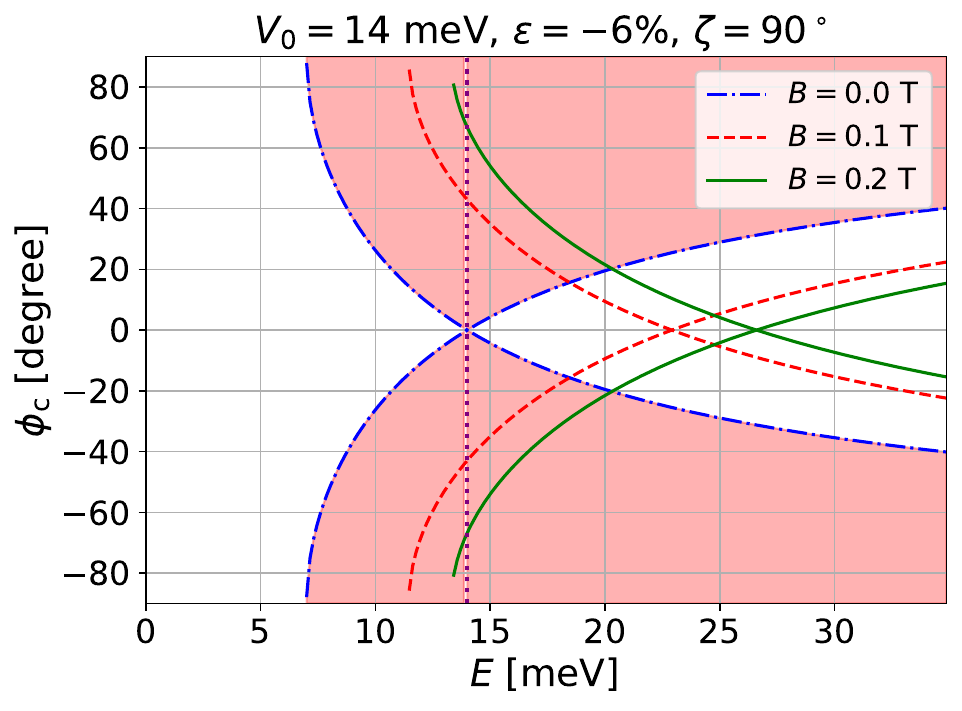}
        \caption{$\epsilon=-6\%$ along the $\mathcal{AC}$ direction}
        \label{fig:3f}
     \end{subfigure}
    \caption{Phase diagram for the wave nature inside a magnetic and electrostatic barrier according to the incidence angle $\phi_{c}$ in \eqref{eq.38} for different deformation configurations along the $\mathcal{ZZ}$ or $\mathcal{AC}$ direction and different magnetic field strengths $B$. In all cases, we set $V_0=14$ meV (vertical purple-dotted line). White regions identified the zone of oscillating waves, while red regions are the zone of evanescent waves.}
    \label{fig:critical_angle}
\end{figure}

Moreover, for certain incidence energies, there exists a critical incidence angle $\phi_{\mathrm{c}}$ at which the $x$ component of the transmitted wave vector $\vec{k}_{1}^{t}$ becomes imaginary, causing the incident plane waves to become evanescent as they propagate inside the barrier. The condition that determines such a critical angle is:
\begin{equation}\label{eq.38}
    \phi_{{\rm c}}=\arcsin\left\vert\frac{v_{y}\left(E-V_{0}-ev_{y}A_{y}(x)\right)}{\sqrt{v_{x}^{2}E^{2}+(v_{y}^{2}-v_{x}^{2})\left(E-V_{0}-ev_{y}A_{y}(x)\right)^{2}}}\right\vert.
\end{equation}
Thus, electrons with an energy $E$ for which there is no critical angle (i.e., the absolute value of the argument in \eqref{eq.38} is greater than 1) cannot undergo total reflection at the barrier and are transmitted as plane waves. For an incident particle whose energy $E$ guarantees that the absolute value of the argument in \eqref{eq.38} is less than 1, it can only be transmitted as a plane wave if its incidence angle satisfies $\vert \phi_{0}\vert <\phi_{{\rm c}}$, where the value of $\phi_{{\rm c}}$ is determined by energy itself (see Fig.~\ref{fig:critical_angle}).

In other words, without a magnetic field, there is a zone of oscillating waves, or classically allowed region, for particles with incident energy $E<V_0/2$ that makes possible resonances and whose area reduces when $1/2<E/V_0<1$. Additionally, there is a zone of evanescent waves, or classically forbidden, that provides the possibility of a tunnel effect via an evanescent wave. These regions are identified as white and red areas, respectively, in Fig.~\ref{fig:3a}, which correspond to the pristine case. The critical angle in Eq. \eqref{eq.38} (blue curve in Fig.~\ref{fig:critical_angle}) delimits the areas of such regions, which can also be modified when a mechanical deformation is applied to the material (see Figs.~\ref{fig:3c} - \ref{fig:3f}). When a magnetic field is applied (red and green curves in Fig.~\ref{fig:critical_angle}), the condition $E<V_0/2$ for oscillating waves is not fulfilled, and the regions described previously are displaced to the right, corresponding to larger energy values.

% Finally, when $E=V_{0}+ev_{y}A_{y}(x)$, we have $\phi_{c}=0$, which means that only particles with normal incidence can be transmitted as plane waves inside the barrier.
%, for given values of the strain $\epsilon$ and the magnetic field strength $B_{0}$

%\section{Numerical results and discussion}\label{sec4}
\subsection{Numerical case: multi-barrier structure}

Let us consider a uniform set of electrostatic potential barriers of height $V_0=14$ meV and $\delta$-magnetic barriers of strength $B=0.1$ T with width $D$ and separation $L$ equal to $\ell_B \approx 81. 1$ nm, as illustrated in Fig.~\ref{fig:Nmbarrier}.

\begin{figure}[htbp]
     \centering
%     \caption*{Transmission Coefficient $T(\theta,E)$}
     \begin{subfigure}{0.45\linewidth}
         \centering
    \includegraphics[width=0.81\linewidth]{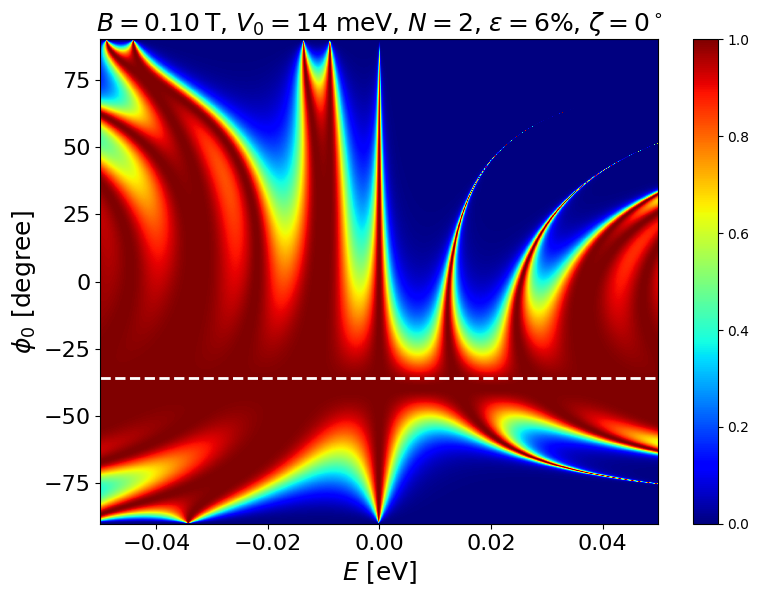}
        \caption{$N=2$, $\zeta=0^\circ$}
        \label{fig:6a}
     \end{subfigure}
     %\hfill
     \begin{subfigure}{0.45\linewidth}
         \centering
    \includegraphics[width=0.81\linewidth]{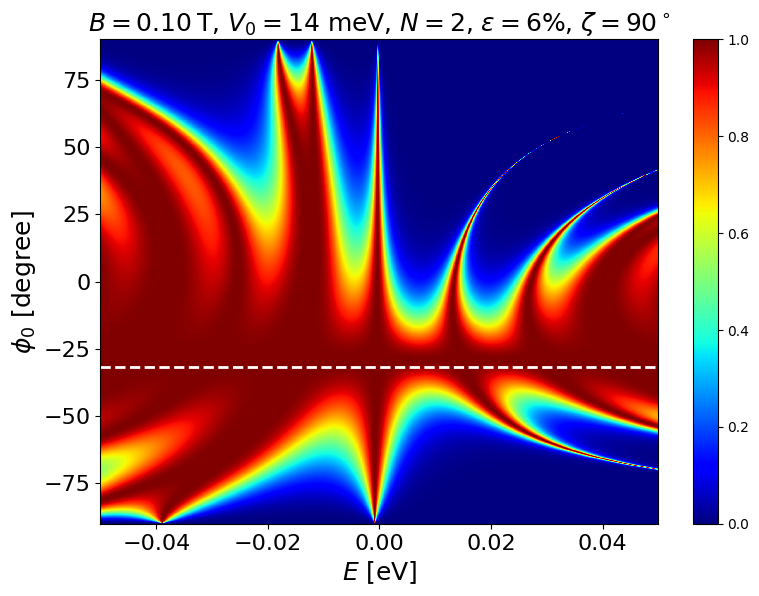}
        \caption{$N=2$, $\zeta=90^\circ$}
        \label{fig:6b}
     \end{subfigure}
     \begin{subfigure}{0.45\linewidth}
         \centering
        \includegraphics[width=0.81\linewidth]{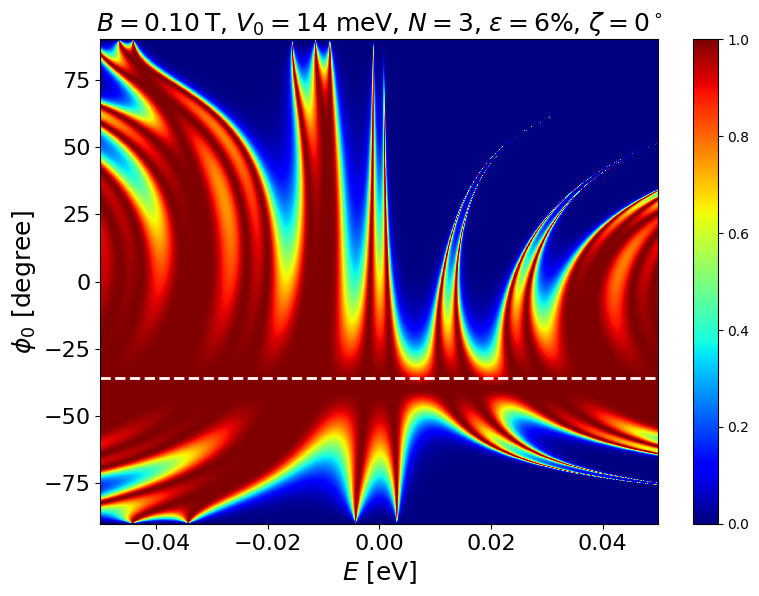}
        \caption{$N=3$, $\zeta=0^\circ$}
        \label{fig:6c}
     \end{subfigure}
     %\hfill
     \begin{subfigure}{0.45\linewidth}
         \centering
        \includegraphics[width=0.81\linewidth]{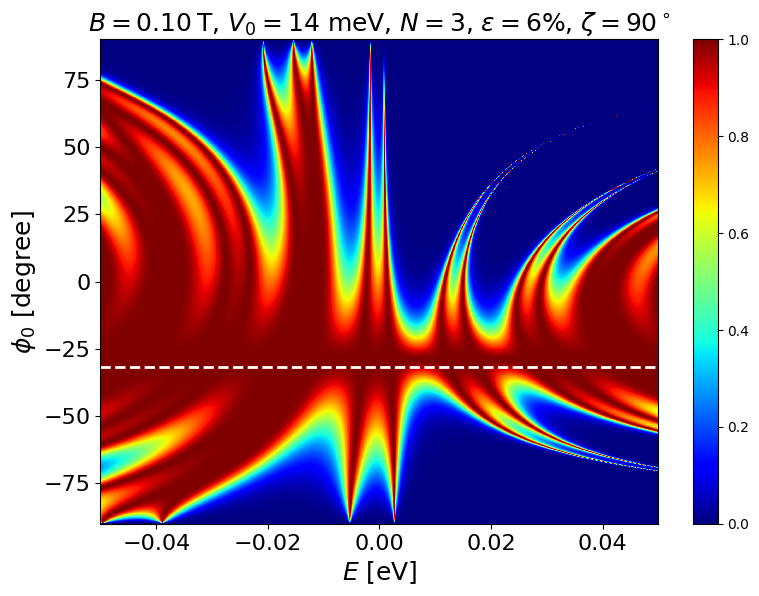}
        \caption{$N=3$, $\zeta=90^\circ$}
        \label{fig:6d}
     \end{subfigure}
     \begin{subfigure}{0.45\linewidth}
         \centering
    \includegraphics[width=0.81\linewidth]{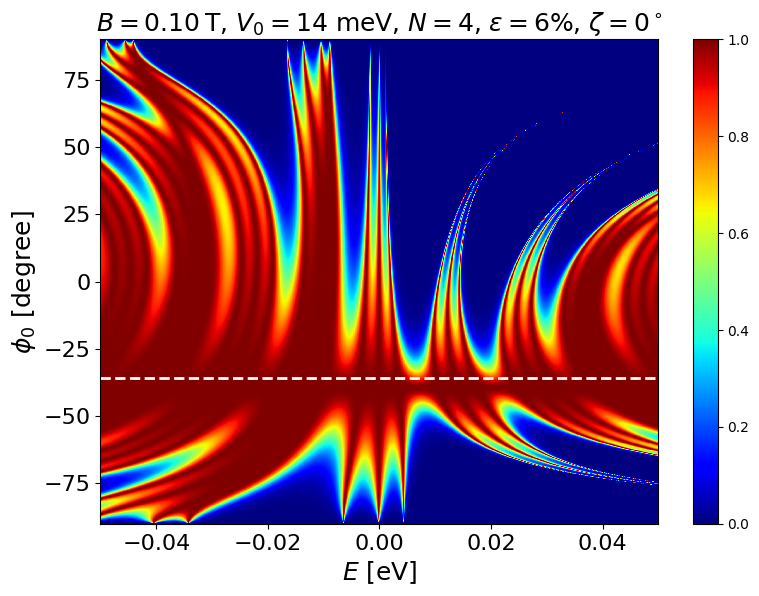}
        \caption{$N=4$, $\zeta=0^\circ$}
        \label{fig:6e}
     \end{subfigure}
     %\hfill
     \begin{subfigure}{0.45\linewidth}
         \centering
    \includegraphics[width=0.81\linewidth]{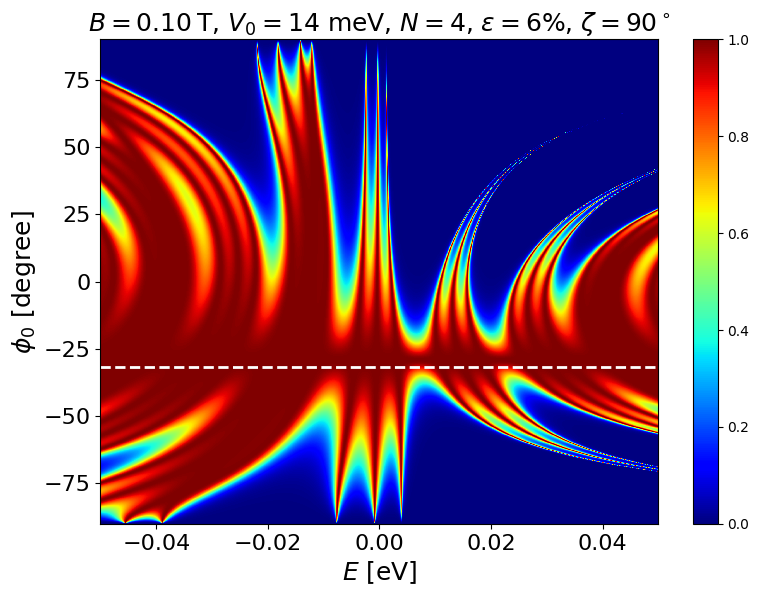}
        \caption{$N=4$, $\zeta=90^\circ$}
        \label{fig:6f}
     \end{subfigure}
     \begin{subfigure}{0.45\linewidth}
         \centering
        \includegraphics[width=0.81\linewidth]{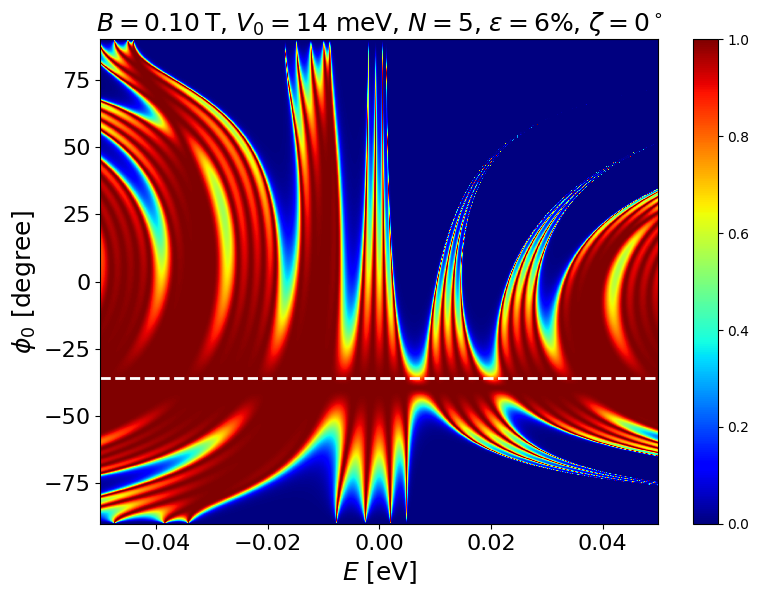}
        \caption{$N=5$, $\zeta=0^\circ$}
        \label{fig:6g}
     \end{subfigure}
     %\hfill
     \begin{subfigure}{0.45\linewidth}
         \centering
        \includegraphics[width=0.81\linewidth]{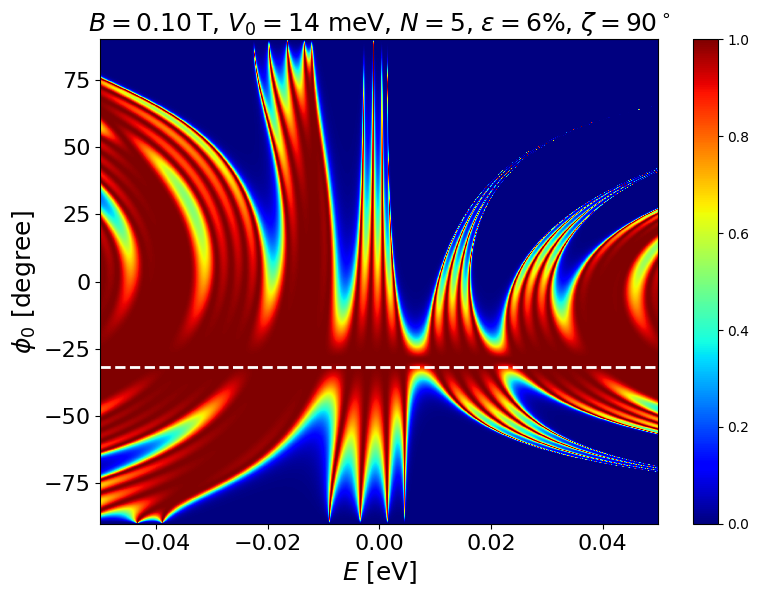}
        \caption{$N=5$, $\zeta=90^\circ$}
        \label{fig:6h}
     \end{subfigure}
    \caption{Electron transmission $T$ in uniaxially strained graphene with $\epsilon = 6\%$, as a function of the Fermi energy $E$ and the incidence angle $\phi_{0}$, for different numbers $N$ of electrostatic and magnetic barriers. The LHS panels correspond to deformations along the $\mathcal{ZZ}$ direction, while the RHS panels correspond to deformations along the $\mathcal{AC}$ direction. In all cases, the parameters are set to $V_0 = 14$ meV, $B = 0.1$ T, and $L=D \approx 81.1$ nm. \textcolor{black}{The dashed white line indicates the anomalous Klein tunnelling direction that occurs for the angle predicted in Eq. \eqref{KTanomalous}.}}
    \label{fig:Nbarrier}
\end{figure}

\begin{figure}[htbp]
     \centering
%     \caption*{Transmission Coefficient $T(\theta,E)$}
     \begin{subfigure}{0.45\linewidth}
         \centering
    \includegraphics[width=0.81\linewidth]{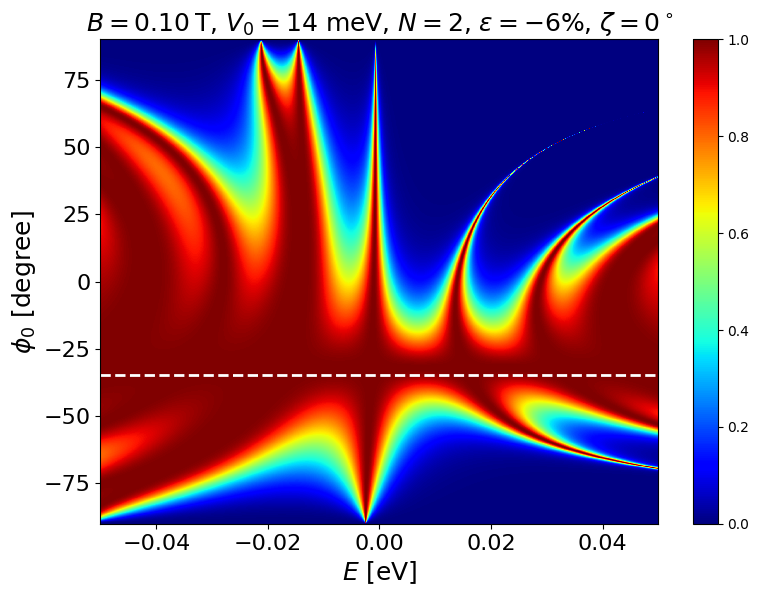}
        \caption{$N=2$, $\zeta=0^\circ$}
        \label{fig:7a}
     \end{subfigure}
     %\hfill
     \begin{subfigure}{0.45\linewidth}
         \centering
    \includegraphics[width=0.81\linewidth]{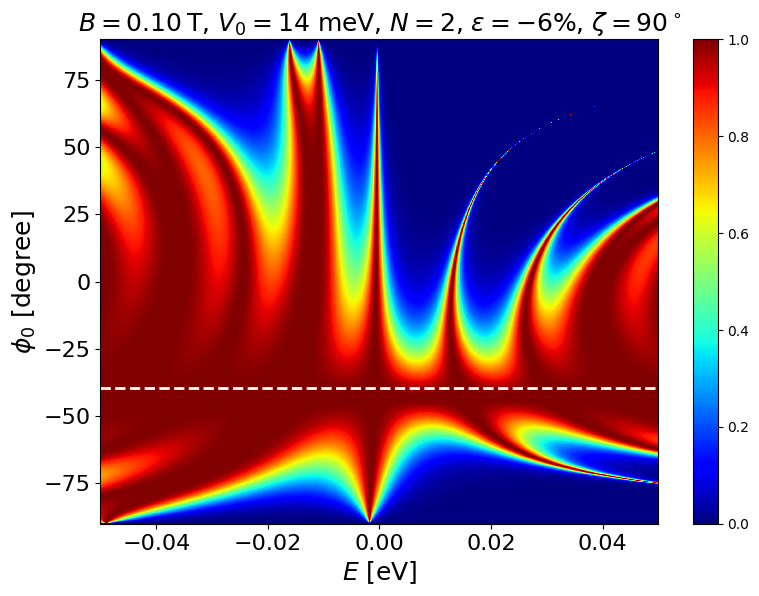}
        \caption{$N=2$, $\zeta=90^\circ$}
        \label{fig:7b}
     \end{subfigure}
     \begin{subfigure}{0.45\linewidth}
         \centering
        \includegraphics[width=0.81\linewidth]{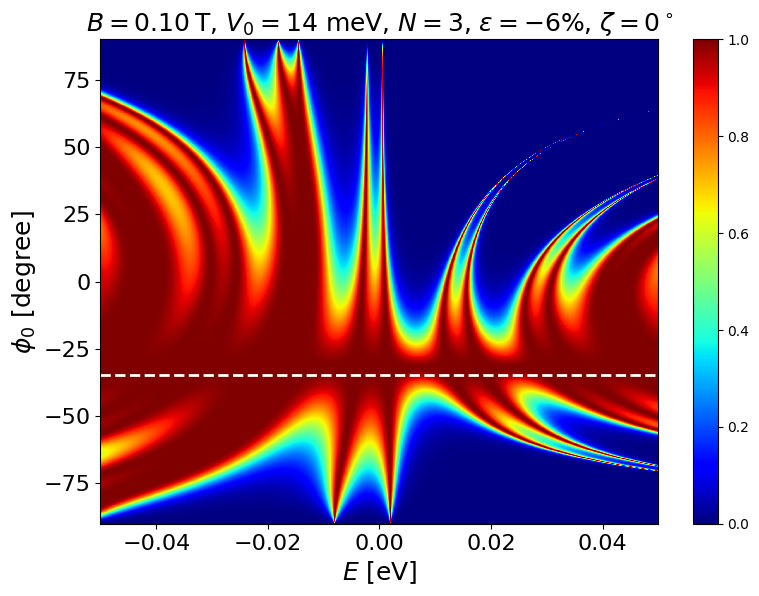}
        \caption{$N=3$, $\zeta=0^\circ$}
        \label{fig:7c}
     \end{subfigure}
     %\hfill
     \begin{subfigure}{0.45\linewidth}
         \centering
        \includegraphics[width=0.81\linewidth]{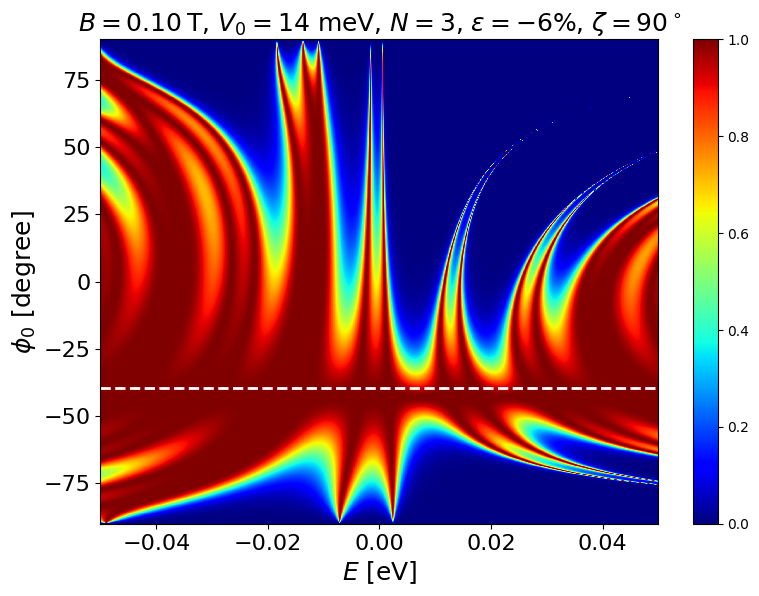}
        \caption{$N=3$, $\zeta=90^\circ$}
        \label{fig:7d}
     \end{subfigure}
     \begin{subfigure}{0.45\linewidth}
         \centering
    \includegraphics[width=0.81\linewidth]{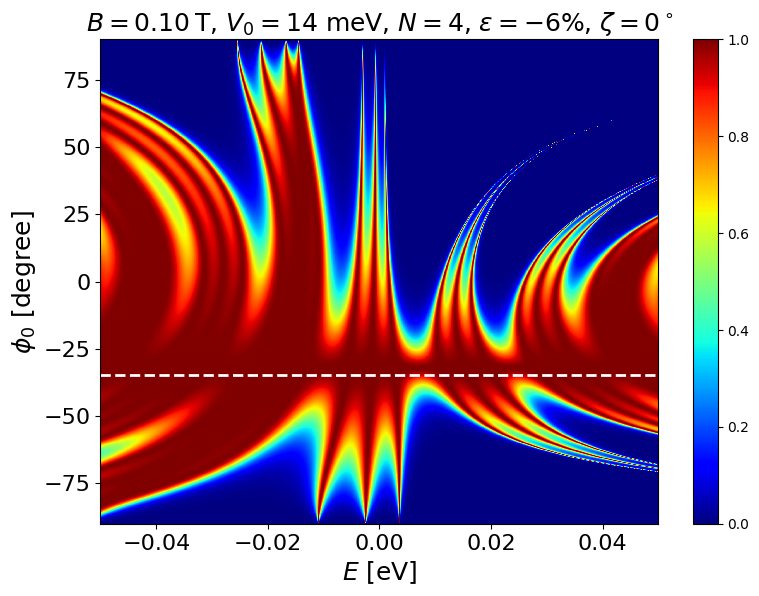}
        \caption{$N=4$, $\zeta=0^\circ$}
        \label{fig:7e}
     \end{subfigure}
     %\hfill
     \begin{subfigure}{0.45\linewidth}
         \centering
    \includegraphics[width=0.81\linewidth]{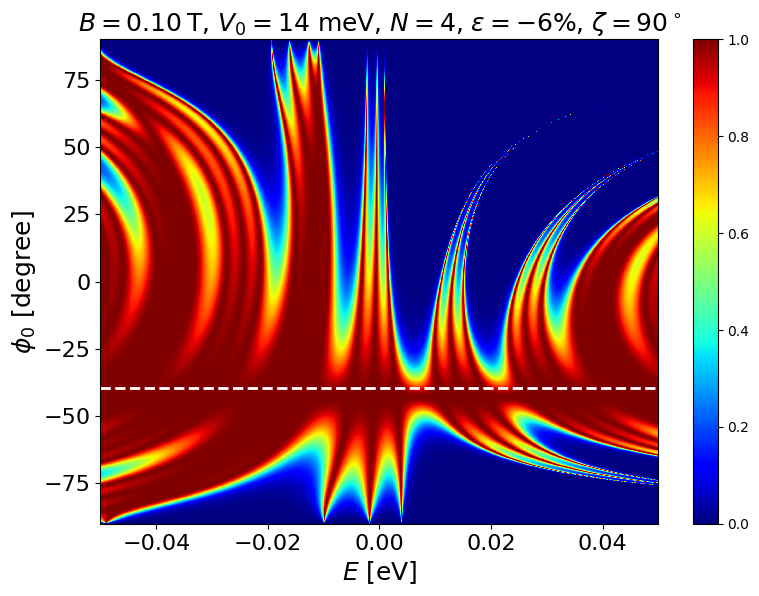}
        \caption{$N=4$, $\zeta=90^\circ$}
        \label{fig:7f}
     \end{subfigure}
     \begin{subfigure}{0.45\linewidth}
         \centering
        \includegraphics[width=0.81\linewidth]{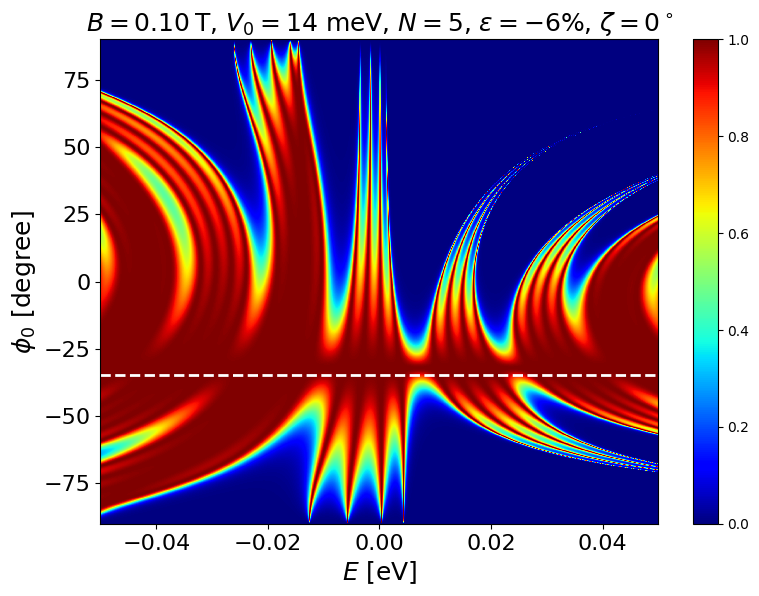}
        \caption{$N=5$, $\zeta=0^\circ$}
        \label{fig:7g}
     \end{subfigure}
     %\hfill
     \begin{subfigure}{0.45\linewidth}
         \centering
        \includegraphics[width=0.81\linewidth]{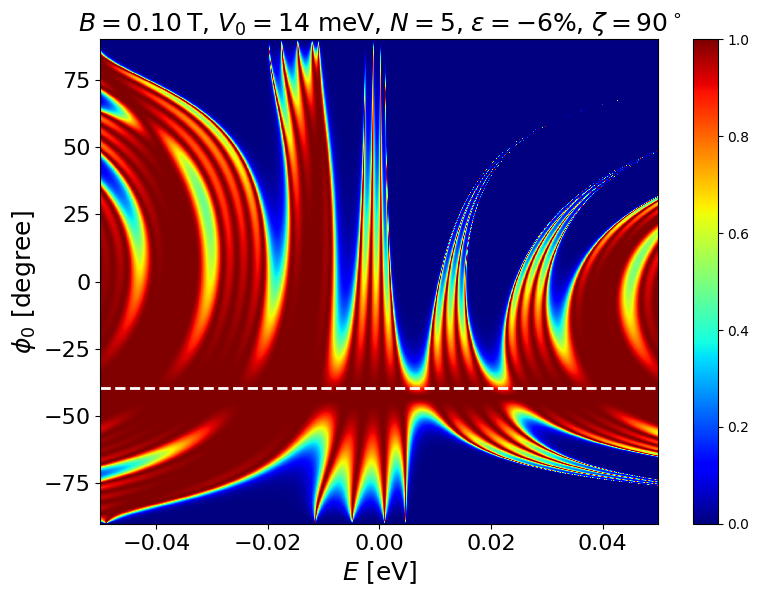}
        \caption{$N=5$, $\zeta=90^\circ$}
        \label{fig:7h}
     \end{subfigure}
    \caption{Electron transmission $T$ in uniaxially strained graphene with $\epsilon = -6\%$, as a function of the Fermi energy $E$ and the incidence angle $\phi_{0}$, for different numbers $N$ of electrostatic and magnetic barriers. The LHS panels correspond to deformations along the $\mathcal{ZZ}$ direction, while the RHS panels correspond to deformations along the $\mathcal{AC}$ direction. In all cases, the parameters are set to $V_0 = 14$ meV, $B = 0.1$ T, and $L=D \approx 81.1$ nm. \textcolor{black}{The dashed white line indicates the anomalous Klein tunnelling direction that occurs for the angle predicted in Eq. \eqref{KTanomalous}.}}
    \label{fig:Nbarrier_compress}
\end{figure}

Here, Figs.~\ref{fig:Nbarrier} and~\ref{fig:Nbarrier_compress} show the behavior of transmission coefficient as a function of the incidence angle $\phi_0$ and Fermi energy $E$ for different values of $N$, considering strained graphene at $\epsilon=6 \%$ (tensile deformation) and $\epsilon=-6 \%$ (compress deformation), respectively, along the $\mathcal{ZZ}$ $(\zeta=0^\circ)$ and $\mathcal{AC}$ $(\zeta=90^\circ)$ direction. In general, for $N=2$ to $N=5$, it can be observed that the number of bands and minigaps increases with the number of barriers. In addition, in all cases studied in this section, it can be appreciated that the behavior described in the single-barrier section is preserved, namely, the absence of a Klein tunnelling at $\phi_0=0^{\circ}$ and the emergence of anomalous Klein tunnelling for different incidence angles.

On the other hand, when we compare the deformation along the $\mathcal{ZZ}$ direction (first column) with that along the $\mathcal{AC}$ direction (second column) in Fig.~\ref{fig:Nbarrier}, we find notable differences between them. For instance, in the first case, anomalous Klein tunnelling appears in the interval {\color{black}$-40^\circ<\phi_{\rm KT}<-35^\circ$, while in the second one, such an effect occurs in the range of $-32^\circ<\phi_{\rm KT}<-27^\circ$}, which is in agreement with the results shown in Figs.~\ref{fig:4a} and~\ref{fig:4b} for a single barrier. Moreover, such behavior persists as the number of barriers increases. In addition, it is clear at this point that, independent of the number of barriers, the incidence angle for which the anomalous Klein tunnelling appears closer to the normal incidence for tensile deformations along the $\mathcal{AC}$ direction. In comparison with the pristine case \cite{Katsnelson2006, Allain2011}, the transmission coefficient suffers less modification when the tensile deformation is applied along $\mathcal{AC}$ than when it occurs along the orthogonal direction, and the material is under the interaction of magnetic barriers. Then, when a compression is applied, as shown in Fig.~\ref{fig:Nbarrier_compress}, the above behaviors exchange with each other, resulting in a minor change in electron transmission $T$ (compared to the pristine case) when such a mechanical deformation is applied along the $\mathcal{ZZ}$ direction. {\color{black}Indeed, according to Eq.~\eqref{KTanomalous}, for the values in Fig.~\ref{fig:Nbarrier_compress}, anomalous Klein tunnelling appears at $\phi_{\rm KT}\approx-34.76^\circ$ for compression deformations along the $\mathcal{ZZ}$ direction, while $\phi_{\rm KT}\approx-39.50^\circ$ when the compression is applied along the $\mathcal{AC}$ direction. Compared to a single-barrier case, this anomalous Klein tunnelling angle persists in the multiple-barrier structure because it is composed of a sequence of identical barriers, which are characterized by the same electrostatic potential height $V_0$ and the vector potential profile $\vec{A}(x)$. Consequently, Eq. \eqref{KTanomalous} remains applicable to a system of multiple identical barriers. This robustness is due to the conservation of pseudospin, which holds equally in regions with and without external potentials. As a result, the electron beam maintains a straight line propagation across the entire multi-barrier structure.}

\begin{figure}[htbp]
     \centering
     \begin{subfigure}{0.45\linewidth}
         \centering
    \includegraphics[width=0.85\linewidth]{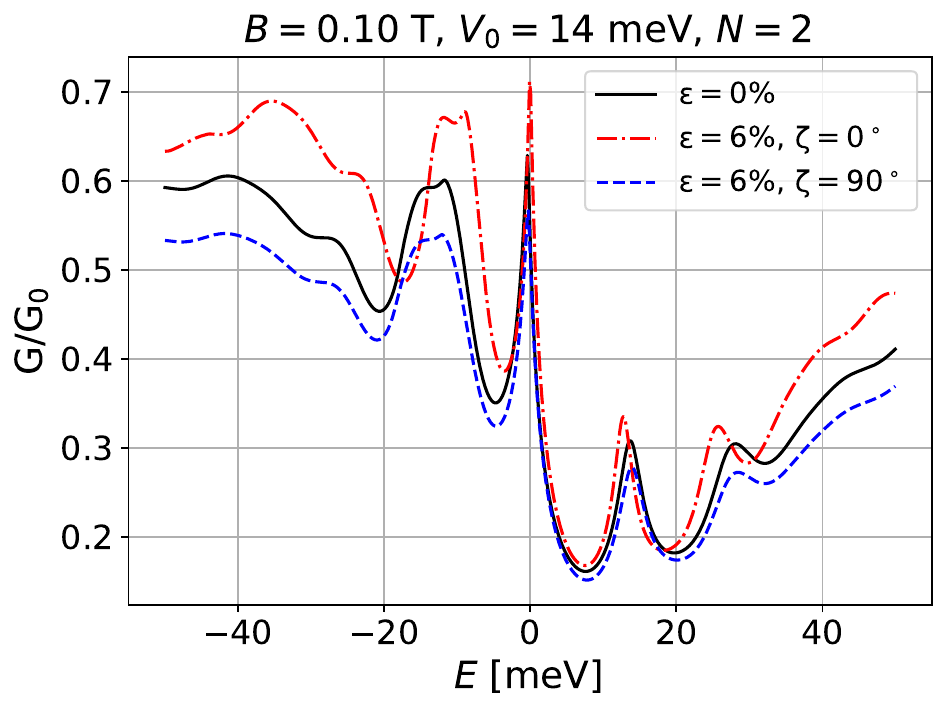}
        \caption{$N=2$}
        \label{fig:8a}
     \end{subfigure}
     %\hfill
    \begin{subfigure}{0.45\linewidth}
         \centering
    \includegraphics[width=0.85\linewidth]{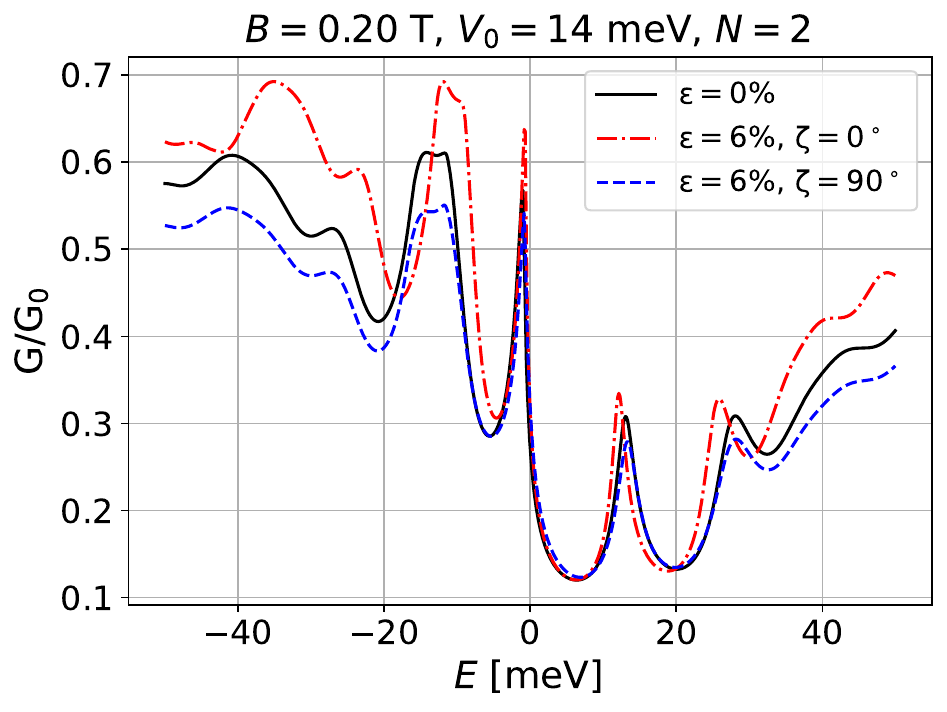}
        \caption{$N=2$}
        \label{fig:8b}
     \end{subfigure}
     \begin{subfigure}{0.45\linewidth}
         \centering
    \includegraphics[width=0.85\linewidth]{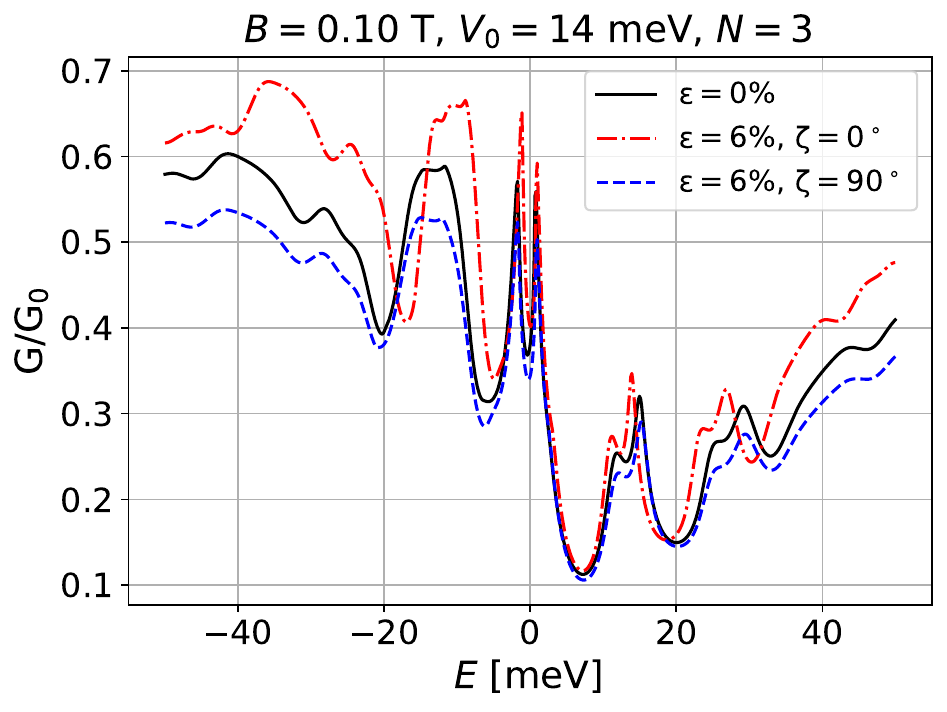}
        \caption{$N=3$}
        \label{fig:8c}
     \end{subfigure}
     \begin{subfigure}{0.45\linewidth}
         \centering
    \includegraphics[width=0.85\linewidth]{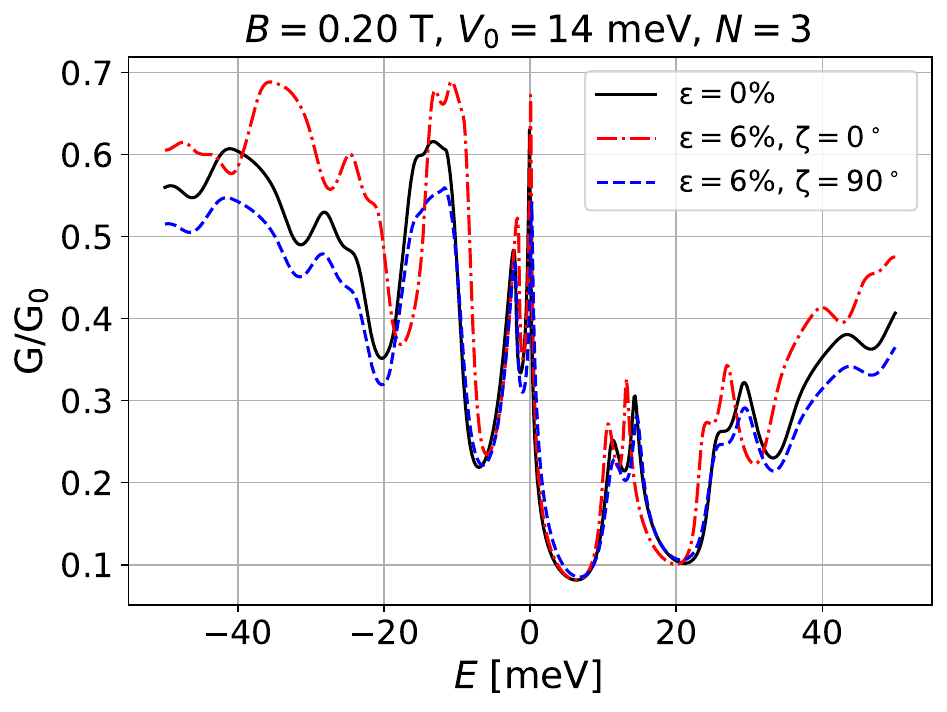}
        \caption{$N=3$}
        \label{fig:8d}
     \end{subfigure}
     \begin{subfigure}{0.45\linewidth}
         \centering
        \includegraphics[width=0.85\linewidth]{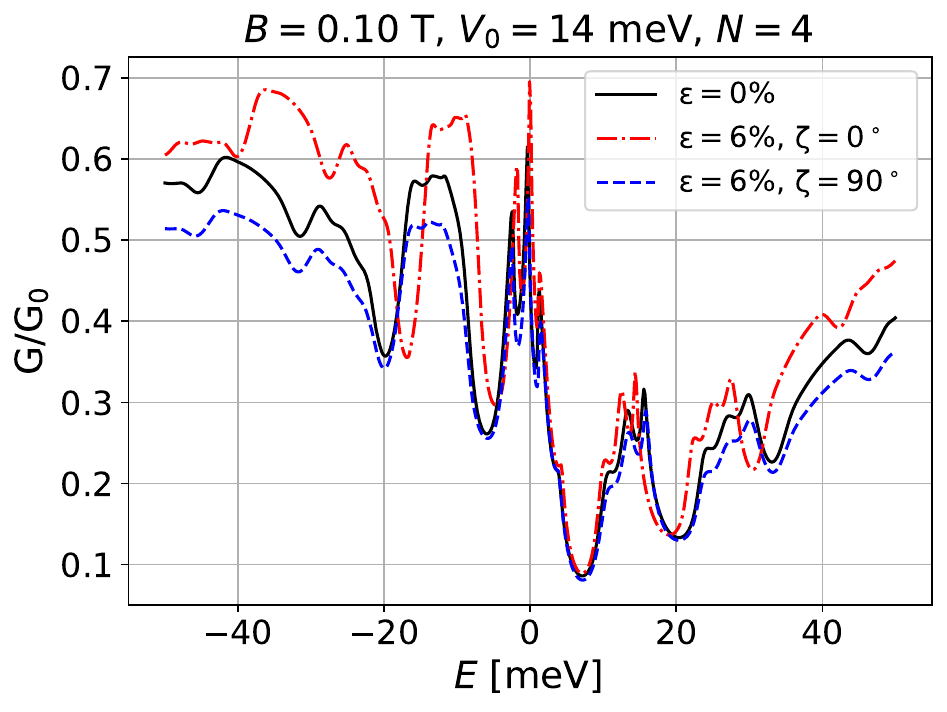}
        \caption{$N=4$}
        \label{fig:8e}
     \end{subfigure}
     %\hfill
     \begin{subfigure}{0.45\linewidth}
         \centering
        \includegraphics[width=0.85\linewidth]{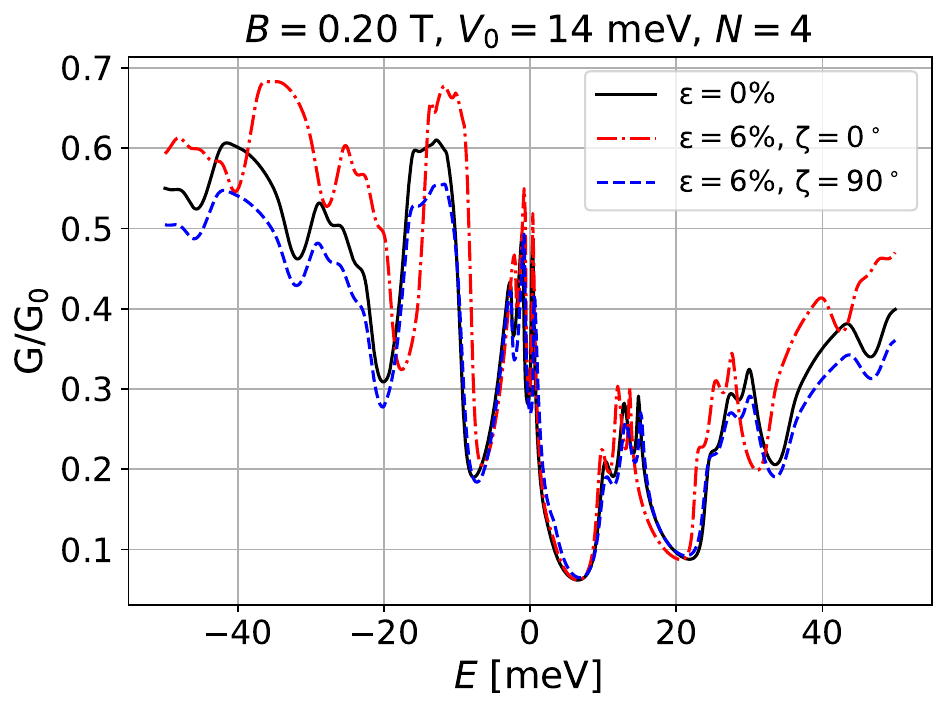}
        \caption{$N=4$}
        \label{fig:8f}
     \end{subfigure}
     \begin{subfigure}{0.45\linewidth}
         \centering
        \includegraphics[width=0.85\linewidth]{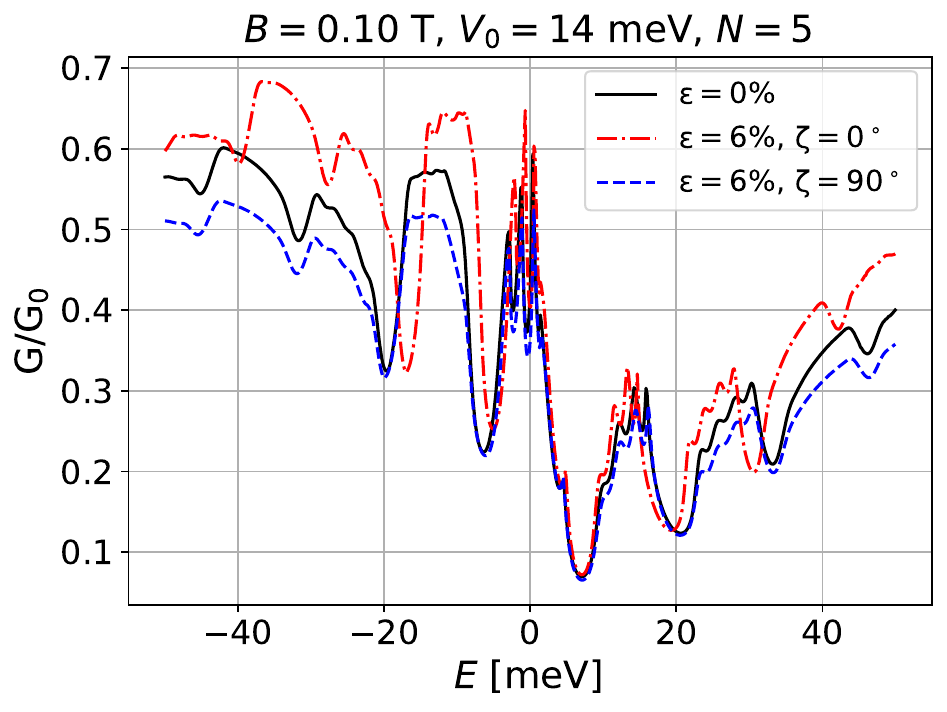}
        \caption{$N=5$}
        \label{fig:8g}
     \end{subfigure}
     \begin{subfigure}{0.45\linewidth}
         \centering
        \includegraphics[width=0.85\linewidth]{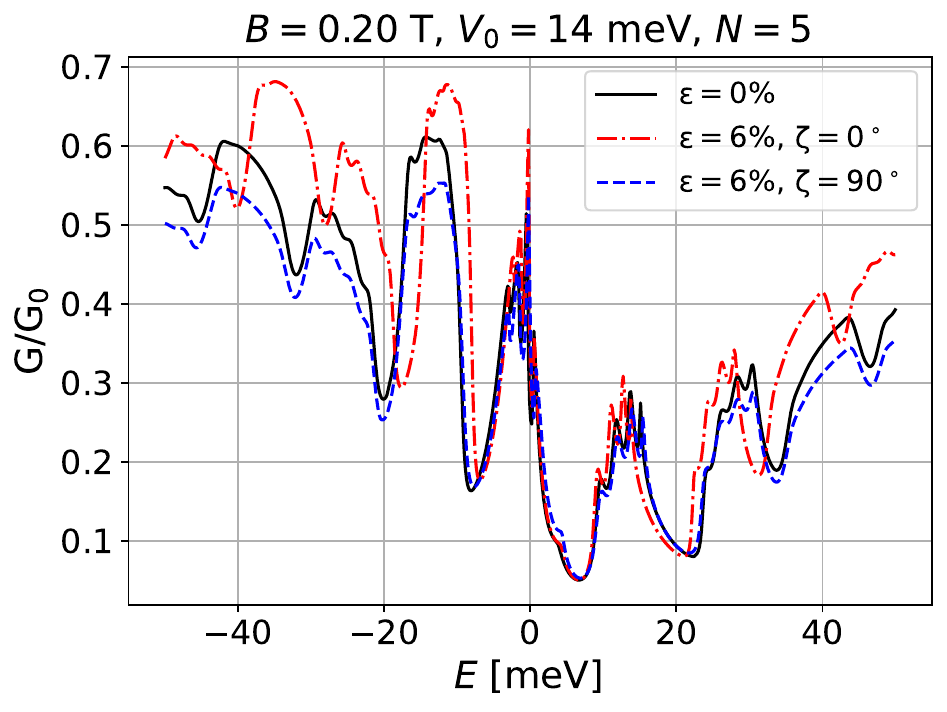}
        \caption{$N=5$}
        \label{fig:8h}
     \end{subfigure}
    \caption{Conductance $G/G_{0}$ as a function of the Fermi level $E$ for different numbers $N \geq 2$ of electrostatic and magnetic barriers with a fixed width $D \approx 81.1$ nm. The solid black, dashed red, and dotted blue curves correspond to pristine graphene and to strained graphene along the $\mathcal{ZZ}$ and $\mathcal{AC}$ directions, respectively. In all cases, the parameters are set to $V_0 = 14$ meV and $B = 0.1$ T ($B=0.2$ T) for LHS (RHS) panels.}
    \label{fig:Nconductance}
\end{figure}

Now, let us focus on the conductance behavior, which is shown in Fig.~\ref{fig:Nconductance}. For $N=2$ to $N=5$, the number of peaks in which the conductance reaches maximum values increases as the number of barriers $N$ does. In all cases and in comparison to the pristine case, conductance tends to increase when a tensile deformation is applied along the $\mathcal{ZZ}$ direction. In contrast, resistance tends to increase when such deformation acts on the orthogonal direction. Similar to the single-barrier case, the conductance exhibits a pronounced drop near $E=0$ meV; however, it begins to increase again once $E>20$ meV, approximately. On the other hand, although the conductance behavior observed in the single-barrier case is not as clear here, since the pristine conductance 
does not lie between the other two cases but intersects with the other conductance curves for given values of the Fermi energy, the general effect of a growing magnetic field strength is to increase resistance. This result can be understood considering that, though the configuration of the magnetic barriers does not allow Landau levels due to the abrupt change in the magnetic field profile, their action is sufficient to alter the dynamics of the charged particles, modifying the constructive interference in each region that the electron traverses.

\section{Conclusions}\label{conclusions}
In this work, through the construction of a transfer matrix framework for a multi-barrier system composed of magnetic and electrostatic potentials, we describe the electron transmission of Dirac fermions in uniaxially strained graphene. This enables a major comprehension of how strain-engineering can be combined with magnetic interactions to modulate electron properties in graphene, for instance.

Our results show that, in general, an anomalous Klein tunnelling arises as a consequence of mechanical deformations applied to the material, as well as a set of $\delta$-function magnetic barriers placed along a spatial direction on the material surface. In contrast to the usual anomalous Klein tunnelling \cite{Betancur2,díazbautista2022extended,Zhou19}, which occurs, for instance, when a mechanical deformation is applied in an oblique direction with respect to the $\mathcal{ZZ}$ and $\mathcal{AC}$ directions \cite{díazbautista2022extended}, here, such tunnelling can also be modulated via $\delta$-function magnetic field barriers. Moreover, conductance is affected by the combined effect of strain and the magnetic field applied.

It is important to remark that although, as Fig.~\ref{fig:BrillouinZone} suggests, tensile deformations along $\mathcal{ZZ}$ direction can be considered as analogues to compress deformations along the $\mathcal{AC}$ direction, and vice versa, as a result of the Poisson ratio of graphene, and this is observed in Figs.~\ref{fig:5a} and \ref{fig:5b} for conductance, when the strength of magnetic barriers is modified, such equivalence is not fulfilled, as shown in Fig.~\ref{fig:5d}. This fact is also obtained when the RHS panels in Fig.~\ref{fig:Nbarrier} are compared with the LHS ones in Fig.~\ref{fig:Nbarrier_compress}, where the minibands and minigaps do not occur for the same Fermi energies, as well as the anomalous Klein tunnelling.

{\color{black} 
Now, while our model provides valuable insights into the combined effects of strain and magnetic barriers on Klein tunnelling, we acknowledge certain approximations that constrain its direct quantitative comparison with experiments. In particular, the continuum Dirac description assumes negligible valley mixing, which is valid for smooth potential variations but breaks down for atomic-scale disorder or sharp defects. Additionally, the $\delta$-function magnetic barriers, while mathematically convenient, neglect Landau level formation and quantum confinement effects that would emerge in realistic finite-width barriers. Extensive theoretical work has investigated the trapping and confinement of Dirac-like electrons by inhomogeneous magnetic fields in graphene~\cite{DeMartino2007b, DEMARTINO2007,Libisch2010,CohnitzLaura2015, Slota2018}. These studies have shown that magnetic field inhomogeneities can create localized states and quantum dot behavior, effects that are not captured by the $\delta$-function model. For a more quantitative description of realistic experimental setups where magnetic barriers have finite width, the model would need to be extended to include the spatial structure of the magnetic field and the resulting formation of edge states and quasi-bound states~\cite{Silvestrov2007}.

Despite its intrinsic limitations, the $\delta$-function barrier model provides valuable physical insight into the interplay between magnetic confinement, strain-induced anisotropy, and Klein tunnelling. It serves as a useful pedagogical tool and offers qualitative guidance for understanding how magnetic field gradients can modulate electronic transport in strained graphene. In literature, it is reported that $\delta$-function magnetic barriers could be achieved by applying ferromagnetic stripes following, for instance, a Kronig-Penney model \cite{Yesilyurt2016,Matulis1994,RamezaniMasir_2009,Yesilyurt2016_2}, and some results indicate that these types of structures enable modulation of electron valley polarization in graphene \cite{Chen2022}.} 
%Future work extending this model to include finite-width barriers with realistic magnetic field profiles would be a natural next step toward more quantitative predictions for experimental systems.}

%Although in literature it is most common to model magnetic barriers through vector potentials proportional to the $x$-coordinate in a finite region $\vert x\vert <D/2$ \cite{DEMARTINO2007,CohnitzLaura2015}, from an experimental perspective, $\delta$-function magnetic barriers have also been considered by applying ferromagnetic stripes following, for instance, a Kronig-Penney model \cite{Yesilyurt2016,RamezaniMasir_2009,Yesilyurt2016_2}. Some results indicate that these types of structures enable modulation of electron valley polarization in graphene \cite{Chen2022}.

In summary, we consider that this study may help deepen understanding of how anisotropy and external magnetic fields together affect electron transmission and conductance in materials like graphene. {\color{black}Future theoretical work incorporating quantum confinement effects due to finite-width barriers, along with experimental studies on strain-engineered graphene devices with controlled magnetic field landscapes, would provide a more complete picture of anomalous Klein tunnelling in two-dimensional Dirac materials.}

\section*{Acknowledgments}
E.D.-B. and A.R. acknowledge financial support from CONAHCYT Project FORDECYT-PRONACES/61533/2020. They also acknowledge support from SIP-IPN under Grant No. 20254000 and CIC-UMSNH under Grant 18236, respectively. Y.B.-O. acknowledges financial support from UNAM-PAPIIT under the Project IA-102125.

\bibliographystyle{ieeetr}
\bibliography{biblio}  %%% Uncomment this line and comment out the ``thebibliography'' section below to use the external .bib file (using bibtex) .





% \end{thebibliography}

\end{document}